%% file: main.tex
\documentclass[10pt,journal,compsoc]{IEEEtran}

\usepackage{subfigure}
\IEEEoverridecommandlockouts
\usepackage{soul}
\usepackage{multirow}
\usepackage{pifont}
\usepackage{amsmath,amssymb,amsfonts}
\usepackage{wasysym}%
\usepackage{xspace}
\usepackage{xurl}
\usepackage{paralist}
\usepackage[hidelinks]{hyperref}
\usepackage{cleveref}
\usepackage{pgfplots}

\pgfplotsset{compat=1.18}

\usepackage{xcolor}
\definecolor{darkblue}{RGB}{0, 0, 139}

\usepackage{subcaption}
\usepackage{colortbl}
\usepackage[english]{babel} 
\usepackage{booktabs}
\usepackage[most]{tcolorbox}
\usepackage{url}
\usepackage[T1]{fontenc}

\input{commands}

\title{Safety Analysis of CPSs Over-the-air Updates: Test Case Prioritization and Selection} 

\usepackage{ifthen}
\usepackage{algorithmic}
\newboolean{showcomments}
\setboolean{showcomments}{true}
\ifthenelse{\boolean{showcomments}}
{\newcommand{\nb}[2]{
  \fcolorbox{black}{yellow}{\bfseries\sffamily\scriptsize#1}
{\sf\small$\blacktriangleright$\textit{#2}$\blacktriangleleft$}
 }

 \usepackage{algorithm}

}
{\newcommand{\nb}[2]{}
 
}
\usepackage{tcolorbox}

\usepackage{longtable}

\newcommand{\ATheNA}{ATheNA\xspace}

\newcommand{\crisp}{CRISP\xspace}
\newcommand{\apdisc}{AP-DISC\xspace}

\newcommand{\otas}{OTA updates\xspace}

\usepackage{tikz}
\usetikzlibrary{fit}
\usepackage{pgfplots}
\pgfplotsset{compat=1.18}
\usepgfplotslibrary{statistics}
\usepackage[switch]{lineno}
\begin{document}


\author{
Andrea~Bombarda\textsuperscript{\textsection},
Luca~Carminati\textsuperscript{\textsection},
Nunzio~Marco~Bisceglia,
Mehrnoosh~Askarpour, 
Sahar~Kokaly,
Ramesh~S,
Marsha~Chechik,
Claudio~Menghi
\thanks{
A.~Bombarda is with the University of Bergamo, Bergamo, Italy - e-mail: \{andrea.bombarda@unibg.it\};
L.~Carminati is with the University of Bergamo, Bergamo, Italy - e-mail: \{l.carminati13@studenti.unibg.it\};
N. M. Bisceglia is with the Gran Sasso Science Institute, L'Aquila, Italy - email: \{nunziomarco.bisceglia@gssi.it\};
Mehrnoosh Askarpour and Sahar Kokaly are with General Motors Canada and McMaster University, Hamilton, Canada - e-mail: \{mehrnoosh.askarpour@gm.com, sahar.kokaly@gm.com\};
Ramesh S is with General Motors, Warren, Michigan, USA - e-mail: \{ramesh.s@gm.com\}
M. Chechik is with the University of Toronto, Toronto, Canada - e-mail: \{chechik@cs.toronto.edu\};
C.~Menghi is with the University of Bergamo, Bergamo, Italy and the McMaster University, Hamilton, Canada - e-mail: \{claudio.menghi@unibg.it\}\\
\textsuperscript{\textsection} These two authors contributed equally.
}
}

\maketitle

\input{sections/abstract}


\input{sections/introduction}

\input{sections/background}

\input{sections/testselectionandprior}

\input{sections/evaluation}

\input{sections/discussion}
\input{sections/related}
\input{sections/conclusions}

\section*{Data Availability}
Our
implementation and benchmark are publicly available for experiment replication~\cite{ReplicationPackage}.


\bibliographystyle{IEEEtran}
\bibliography{bibliography}

\end{document}

%% file: commands.tex
\definecolor{keywordcolor}{RGB}{127,0,85}

\definecolor{background}{rgb}{0.94,0.95,0.96}

\definecolor{preconditionColor}{HTML}{e6ffe0}

\definecolor{postconditionColor}{HTML}{ffffcc}

\newcommand{\ota}{\textsc{OTA}\xspace}

\newcommand{\simulink}{Simulink\xspace}
\newcommand{\matlab}{MATLAB\xspace}
\newcommand{\Simulink}{Simulink\textsuperscript{\tiny\textregistered}\xspace}

\newcommand{\SLTest}{Simulink\textsuperscript{\tiny\textregistered} Test\textsuperscript{\tiny TM}\xspace}

\definecolor{myBlue}{RGB}{0,133,255}

\definecolor{myOrange}{rgb}{1,0.5,0}

\usepackage[framemethod=TikZ]{mdframed}
\mdfsetup{nobreak=false}
\newenvironment{Answer}[1][]{%
  \ifstrempty{#1}%
  {\mdfsetup{%
    frametitle={%
      \tikz[baseline=(current bounding box.east),outer sep=0pt]
      \node[line width=0pt,anchor=east,rectangle,draw=white,fill=white]
    ;}}
  }%
  {\mdfsetup{%
    frametitle={%
      \tikz[baseline=(current bounding box.east),outer sep=0pt]
      \node[anchor=east,rectangle,draw=white,fill=white]
    {\strut #1};}}%
  }%
  \mdfsetup{innertopmargin=-5pt,linecolor=black,%
            linewidth=0.5pt,topline=true,%
            frametitleaboveskip=\dimexpr-\ht\strutbox\relax,skipabove=\topskip,skipbelow=\topskip}
  \begin{mdframed}[nobreak=false]\relax
  }{\end{mdframed}}

\newtcolorbox{mybox}{colback=red!5!white,colframe=red!75!black}

\newcommand{\Hecate}{\textsc{HECATE}\xspace}

%% file: sections/abstract.tex
\begin{abstract}
Over-the-air (OTA) updates for Cyber-Physical Systems (CPS) allow engineers to modify their software wirelessly.
In the automotive domain, for instance, they enable manufacturers to update software without physical access to the vehicle.
However, OTA updates can generate a large number of possible software configurations. 
For example, end users may decide whether or not to install any given OTA updates.
The combinatorial explosion of these configurations hampers software testing, a standard practice for detecting errors in CPS.
To overcome this limitation, we propose \crisp (CRIticality-based Selection and Prioritization),
a novel framework supporting test case prioritization and selection in the context of OTA updates.
Our framework leverages test case robustness, i.e., how far they are from violating the system requirements, to drive test case prioritization and selection. 
It enables engineers to rank test case executions from most to least critical, and to select only the critical test cases relevant to a given OTA update.
We evaluated our approach using a representative software product line consisting of seven different vehicles, 22 driving scenarios, and two software components.
We considered 224 test cases, requiring 59 hours of execution time.
We assessed the effectiveness of \crisp across five OTA updates.
We compared our solution with (a)~a random test case prioritization and selection algorithm, and (b)~\apdisc, identified as the state-of-the-art solution in a recent empirical study on the subject.
Our results show that \crisp outperforms our baselines in finding failure-revealing test cases, and that our selection procedure can be configured to find reasonable and practical tradeoffs among accuracy (percentage of failure-revealing test cases identified) and computational time (time required to find them).
\crisp allows for saving up to $\approx$4 hours with an accuracy of $95.6\%$ for some OTA updates and, on average, up to $\approx$3.3 hours with an accuracy of $78.8\%$ across all updates.

\end{abstract}

%% file: sections/introduction.tex
\section{Introduction} 
\label{sec:introduction}
Cyber-physical systems (CPS) consist of multiple software (cyber) components~\cite{5995279,liu2020survey,zeng2020survey} acting on their physical environment to provide the desired functionality, combining 
cyber and physical capabilities~\cite{platzer2018logical}.
Considering these capabilities, software failures of CPS can have catastrophic consequences and lead to significant economic losses~\cite{TeslaCrash,Uber}. 
Therefore, engineers need to ensure that hazards are sufficiently mitigated and systems are thoroughly tested~\cite{ericson2015hazard,smith2020safety} to guarantee the system's safety and prevent failures. 

The presence of many components, often developed by third-party companies, hampers hazard mitigation~\cite{el2023analysis,bisceglia2026safety}.
This activity is further complicated by the presence of software updates deployed via over-the-air updates (\ota)~\cite{white2012making}.
\ota updates are becoming the standard practice to update CPS software.
For example, \ota updates are extensively used by automotive manufacturers to modify the vehicle entertainment and navigation system or the software executed by the electronic control units~\cite{kathiresh2021automotive}.
They offer a convenient solution for updating the vehicle's software since customers do not have to travel to the dealership. 
While providing remarkable benefits, \ota updates are challenging software engineering activities~\cite{raeisdanaei2024evaluation,annable2024comprehensive}: the frequent changes and updates in software components can lead to many software configurations that are impossible to analyze manually without automated support.
Consequently, OTA can lead to safety breaches.
For example, Jeep recently sent out an OTA software update that led vehicles to lose drive power~\cite{JeepRecall}, and Tesla issued an OTA update that often disabled Autopilot~\cite{TeslaRecall}.

Software testing is the primary means of preventing such software defects.
It is used in many domains, including real-time, embedded, and safety-critical systems~\cite{Shaukat2010}.
Testing typically feeds some (automatically generated or manually defined) inputs to the software and analyzes its output.
\otas over a fleet of vehicles occurring over a course of time give rise to a multiplicity of software configuration, each of which need to be tested separately. 
Executing the test cases for all the software configurations is impractical and time-consuming, as confirmed by industrial evidence. 
BMW reports delivering OTA software updates to more than 10 million vehicles across multiple brands, model series, and drive types~\cite{BMW}.
Similarly, Volkswagen regularly performs OTA updates for all ID models~\cite{Volkswagen}.
Industrial OTA platforms explicitly identify multi-ECU updates, regional variant complexity, and configuration management as key engineering challenges~\cite{SRMTech}. 
Since modern vehicles and industrial systems may contain between 70 and 150 ECUs, validating OTA updates across all possible configurations is not always possible. 
Therefore, software engineers need techniques for effective prioritization and selection of test cases.

This paper presents \crisp (CRIticality-based Selection and Prioritization), 
a novel framework which reuses requirement-margin (robustness) information computed before an OTA update to guide regression test prioritization and configurable test selection after the update.
Instead of recomputing criticality after each software update, \crisp leverages robustness values from baseline testing as a reusable signal to prioritize and reduce regression testing effort.
Our work is driven by two research hypotheses:
\begin{itemize}
    \item \emph{Hypothesis~1}\label{hyp1} - 
     For OTA updates that preserve the relationship between baseline criticality and post-update behavior, re-executing test cases in descending order of baseline criticality enables earlier detection of failure-revealing test cases.
    \item \emph{Hypothesis~2}\label{hyp2} - For the test case selection following an OTA update, running a single configuration from a set of behaviorally equivalent options improves testing efficiency, i.e., reduces overall execution time, with minimal impact on effectiveness, i.e., capability of detecting the failure-revealing test cases.
\end{itemize}

To validate our research hypotheses, we implemented \crisp in the \simulink~\cite{Simulink} environment, a widely used tool~\cite{elberzhager2013analysis,DBLP:journals/sosym/LiebelMTLH18} employed across industries, such as medical, avionics, and automotive~\cite{dajsuren2013simulink,boll2021characteristics}.
We considered test cases developed using the support provided by \SLTest~\cite{SimulinkTest}, i.e., specified via the \emph{Test Sequence} and \emph{Test Assessment} blocks. 
\emph{Test Sequence} and \emph{Test Assessment} blocks let users construct complex test inputs~\cite{TestSequence} and specify the procedure to check the requirements of interest~\cite{TestAssessment}. 
To assess the criticality of test cases across different inputs, we rely on \Hecate~\cite{10374027}.
It is an academic tool used to perform Search-based Testing of \Simulink models. 
Starting from a Test Assessment block, it generates a metric that measures how far the test case is from violating (or satisfying) the requirements. 
\Hecate has been assessed on a large variety of models, including a cruise controller for an industrial simulator~\cite{Formica2023}, a controller for e-Bikes~\cite{marzella2025test}, models from an international competition on falsification-based testing tools~\cite{DBLP:conf/arch/ErnstABCDFFG0KM21}, from the web (\cite{simulinkpacemaker,TestSequence, benchmarkHPS,benchmarkST,benchmarkFS,benchmarkTL}), a model developed by Toyota~\cite{jin2014powertrain}, and models from a Lockheed Martin benchmark~\cite{mavridou2020ten,benchmarkLM}.
\crisp is implemented as a software prototype within \Hecate, offering a novel way to leverage the robustness metric of \Hecate to guide test case prioritization and selection in line with our research hypothesis. Our prototype offers two execution modes that respectively support the prioritization and selection of the test cases. 

To assess the effectiveness of \crisp, we considered a representative software product line (SPL) consisting of seven different vehicles, 22 driving scenarios, and two software components (an Adaptive Cruise Controller --- ACC and a Lane Keeping Assist --- LKA).
Our software product line is associated with 224 test cases. 
Running all of these test cases requires 59 hours.
We evaluated how capable our test case prioritization solution is in finding failure-revealing test cases (RQ1) and how effective our test case selection solution is in extracting failure-revealing test cases (RQ2).
For RQ1, we compared our test case prioritization solution with two baselines: Random Ordering for test case execution and \apdisc (Anti-patterns based on Discontinuity)~\cite{Arrieta2026}.
Our empirical evaluation also characterizes the conditions under which \crisp remains effective, showing that the approach is most beneficial for incremental OTA updates and SPLs exhibiting limited behavioral variability across configurations.
For RQ2, we assessed our solution's ability in finding failure-revealing test cases and the time required to do so depending on the chosen threshold.
Our solution effectively achieves practical trade-offs between accuracy (percentage of failure-revealing test cases identified) and computational time (time required to find them).
More specifically, \crisp allows for saving up to $\approx$4 hours with an accuracy of $95.6\%$ for some OTA updates and, on average, up to $\approx$3.3 hours with an accuracy of $78.8\%$ across all updates.

To summarize, our contributions are as follows:
\begin{itemize}
    \item A representative exemplar of an automotive SPL obtained by considering seven different vehicles, 22 driving scenarios, and two software components (\Cref{sec:background});
    \item A test suite for the exemplar consisting of $224$ test cases and five different OTA updates targeting two software components: the Lane Keeping Assistant (LKA) and the Adaptive Cruise Controller (ACC)  (\Cref{sec:background}); 
    \item \crisp, a 
    a novel regression-testing technique that reuses requirement-margin (robustness) information computed during baseline testing as a reusable signal for post-update test prioritization and configurable test selection, together with an implementation targeting \matlab/\Simulink (\Cref{sec:method});
    \item An extensive empirical evaluation on our exemplar,
    considering both our test case prioritization and selection techniques, and comparing the effectiveness of our solution with two baselines from the literature that rely on well-established metrics (\Cref{sec:evaluation});
    \item An empirical characterization of the applicability conditions of \crisp, showing that the proposed approach is most effective when OTA updates preserve the relationship between baseline criticality and post-update behavior, and when configuration variability induces limited behavioral differences across configurations (\Cref{sec:discussion});
    \item A thorough reflection on our results and lessons learned, and a discussion of the threats to validity (\Cref{sec:discussion}); 
\end{itemize}

This paper is structured as follows.
\Cref{sec:background} introduces our motivating example and defines our problem.
\Cref{sec:method} outlines our solution.
\Cref{sec:evaluation} evaluates our contribution.
\Cref{sec:discussion} discusses our results.
\Cref{sec:related} discusses related work.
\Cref{sec:conclusion} presents our conclusions.

%% file: sections/background.tex
\begin{figure*}
\subfigure[High-Level View.]{
    \includegraphics[width=\columnwidth]{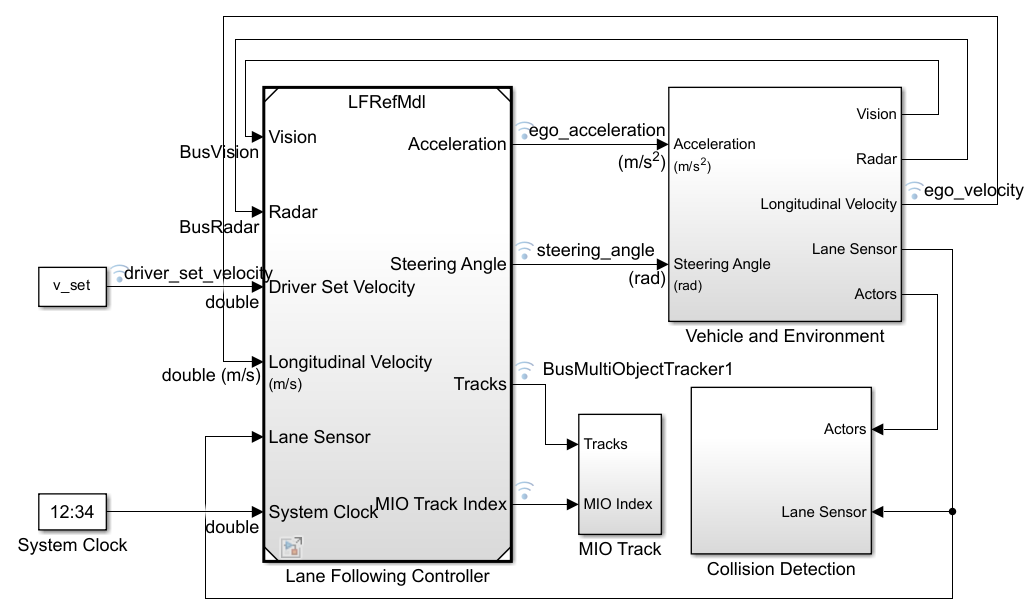}
    \label{fig:SimulinkModel}
}
\subfigure[ACC and LKA components.]{
    \includegraphics[width=\columnwidth]{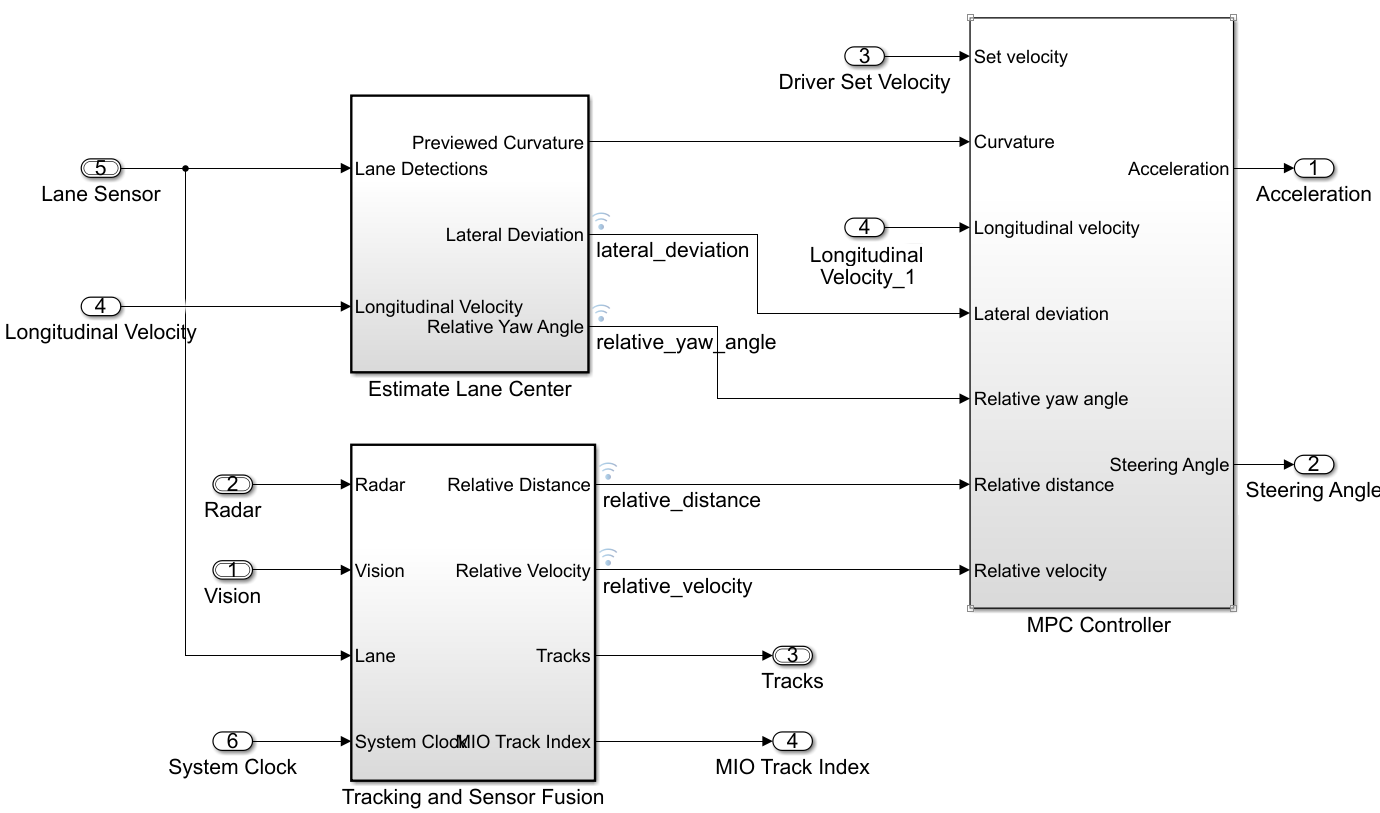}
\label{fig:LaneFollowingwithSpacingControl}
    }
\caption{An excerpt of The \Simulink model of our automotive example.}
\label{fig:Model}
\end{figure*}

\section{Automotive Example and Problem Definition} \label{sec:background}
Our motivating example is a representative model from the automotive domain.
 We describe our model (\Cref{sec:SimulinkModel}), the corresponding software product line (\Cref{sec:SPL}), and the problem of supporting OTA updates (\Cref{sec:ProblemDefinitionOTAs}).

\subsection{The Case Study Simulink Model}
\label{sec:SimulinkModel}
Our motivating example concerns the Advanced Driver-Assistance System (ADAS) of vehicles modeled in \matlab \Simulink, based on a model originally provided by MathWorks~\cite{OriginalModel}.
\Simulink~\cite{Simulink} is widely used to design software~\cite{elberzhager2013analysis,DBLP:journals/sosym/LiebelMTLH18}, and it is used in many industries, such as medical, avionics, and automotive~\cite{dajsuren2013simulink,boll2021characteristics}.
\Cref{fig:Model} shows an excerpt of the \Simulink model of our automotive example.

\Cref{fig:SimulinkModel} shows a high-level view of the model containing its main components. The model comprises three main components: the \emph{Lane Following Controller}, the \emph{Vehicle and Environment}, and the \emph{Collision Detection}.  The \emph{Lane Following Controller} controls the acceleration and the steering angle of the vehicle during the simulation. 
The \emph{Lane Following Controller} monitors the velocity set by the driver (\emph{Driver Set Velocity}), information from the vision (\emph{Vision}) and from the radar (\emph{Radar}), the longitudinal velocity of the vehicle (\emph{Longitudinal Velocity}), and the system clock (\emph{System Clock}).
It acts on the acceleration (\emph{Acceleration}) and on the steering angle (\emph{Steering Angle}), and provides as output the set of objects, i.e., cars, surrounding the ego vehicle (\emph{Tracks}) and the index of the most important object (\emph{MIO Track Index}), i.e., the vehicle that precedes the ego vehicle.

The \emph{Tracks} signal, together with the \emph{MIO Tracks Index} signal, is then used to generate the 2D visualization of the simulation through the Bird's-Eye Scope.

\begin{figure*}
\subfigure[Vehicle view.]{
    \includegraphics[width=\columnwidth]{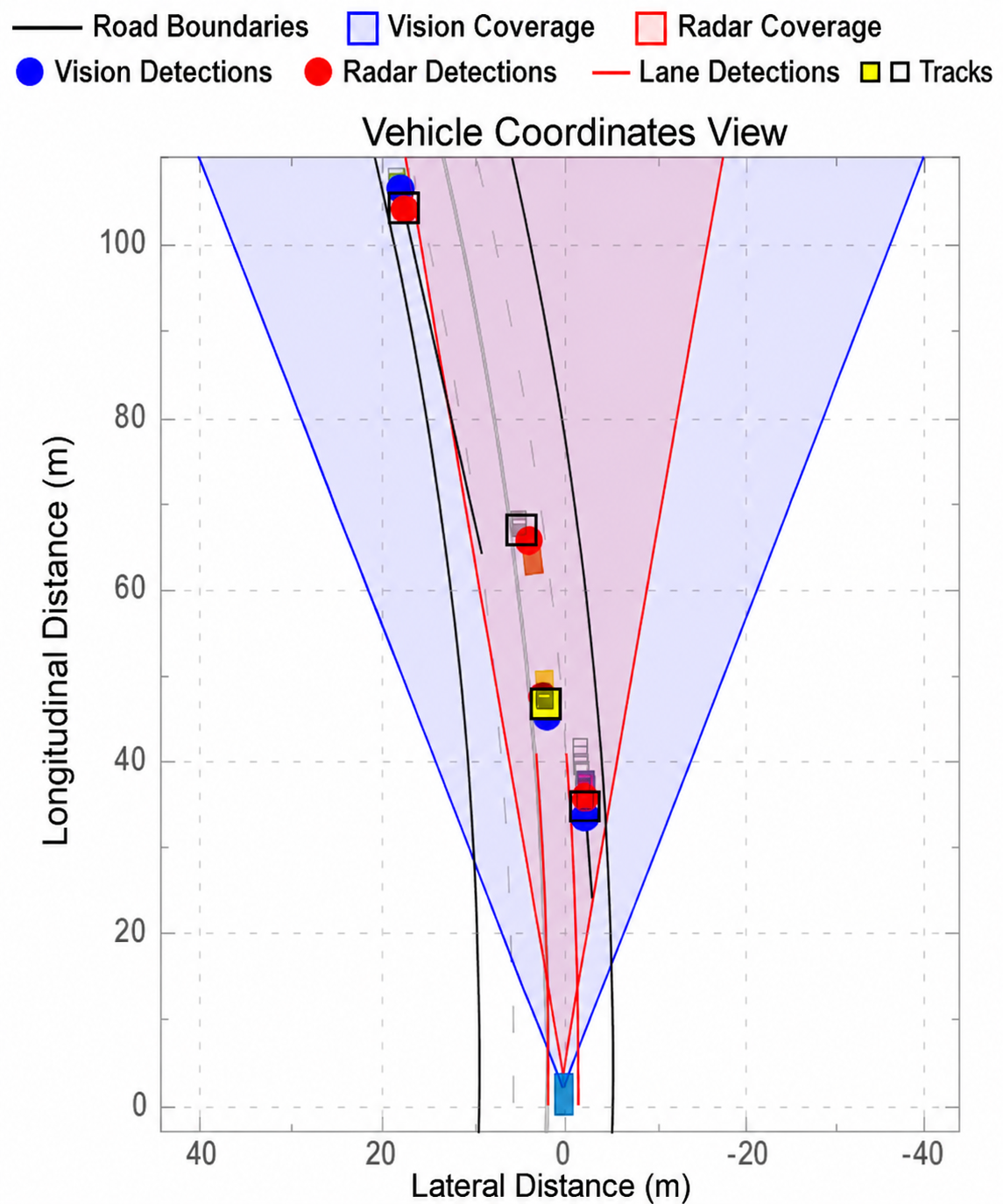}
    \label{fig:Bird's-Eye Scope Vehicle coordinates view}
}
\subfigure[World view.]{
    \includegraphics[width=\columnwidth]{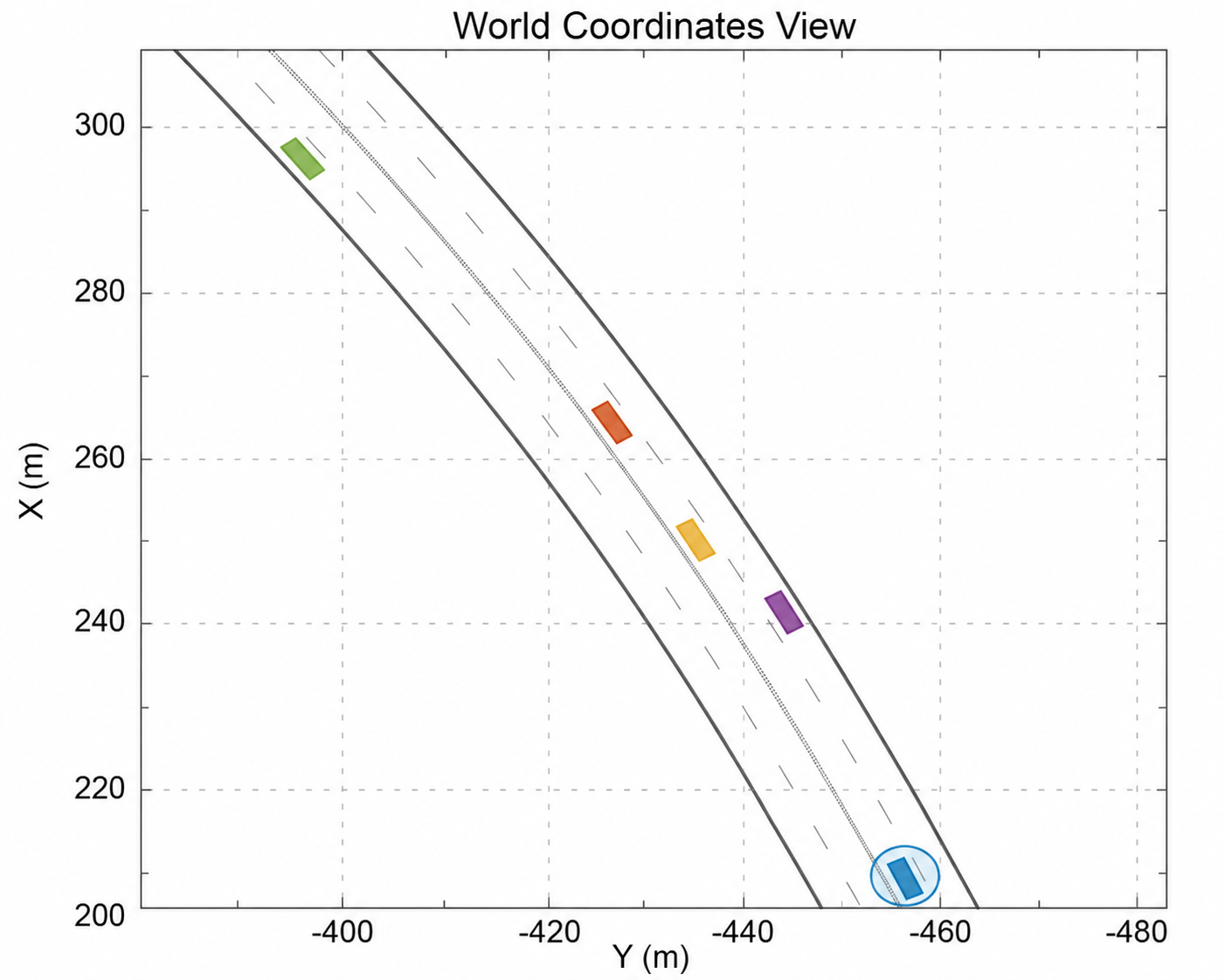}
\label{fig:Bird's-Eye Scope World coordinates view}
    }
\caption{Bird's-Eye Scope view.}
\label{fig:Bird's-Eye Scope}
\end{figure*}


\Cref{fig:LaneFollowingwithSpacingControl} shows the \emph{Lane Following Controller} which contains the software controller implementing our ADAS functionalities: the LKA and ACC. 
The \emph{Estimate Lane Center} block is used to estimate the position of the lane center, to be used by the \emph{Lane Following Controller}. 
It uses information from the \emph{Lane Detections} and the \emph{Longitudinal Velocity} and provides as output the \emph{Previewed Curvature} of the road, its \emph{Lateral Deviation} and \emph{Relative Yaw Angle}.
The \emph{Tracking and Sensor Fusion} is used to estimate relevant information from the \emph{Radar} and \emph{Vision Components}, as well as the \emph{Lane} sensor.
These sources, together with the \emph{System Clock}, the set of objects surrounding the ego vehicle (\emph{Tracks}), and the index of the most important object in that set (\emph{MIO Tracks Index}), allow the system to estimate the \emph{Relative Distance} from the front vehicle and its \emph{Relative Velocity}.
Finally, the \emph{MPC Controller} implements the actual ACC and LKA functionalities.
Based on the \emph{Set velocity} required by the driver, on the \emph{Curvature}, \emph{Lateral deviation}, and the \emph{Relative yaw angle} of the lane cente, on the \emph{Longitudinal velocity}, and on the \emph{Relative distance} and \emph{Relative velocity} of the front vehicle, the \emph{MPC Controller} block provides the required \emph{Acceleration} and \emph{Steering Angle} to keep the lane (LKA) and maintain the required velocity (ACC).

\begin{figure*}
\subfigure[A4 Bergamo satellite view.]{
    \includegraphics[width=\columnwidth]{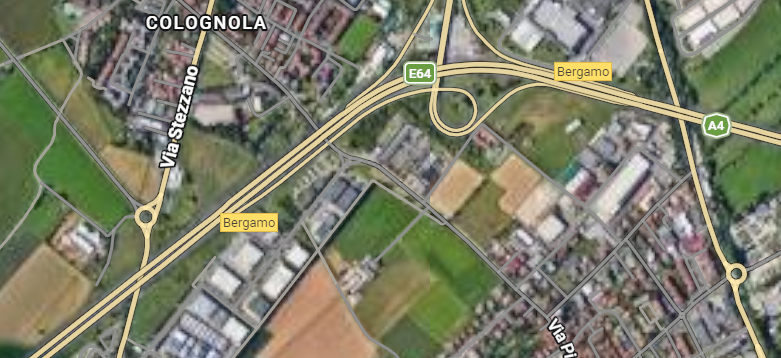}
    \label{fig:A4BergamoSatellite}    
    }
\subfigure[Driving scenario replica.]{
    \includegraphics[width=\columnwidth]{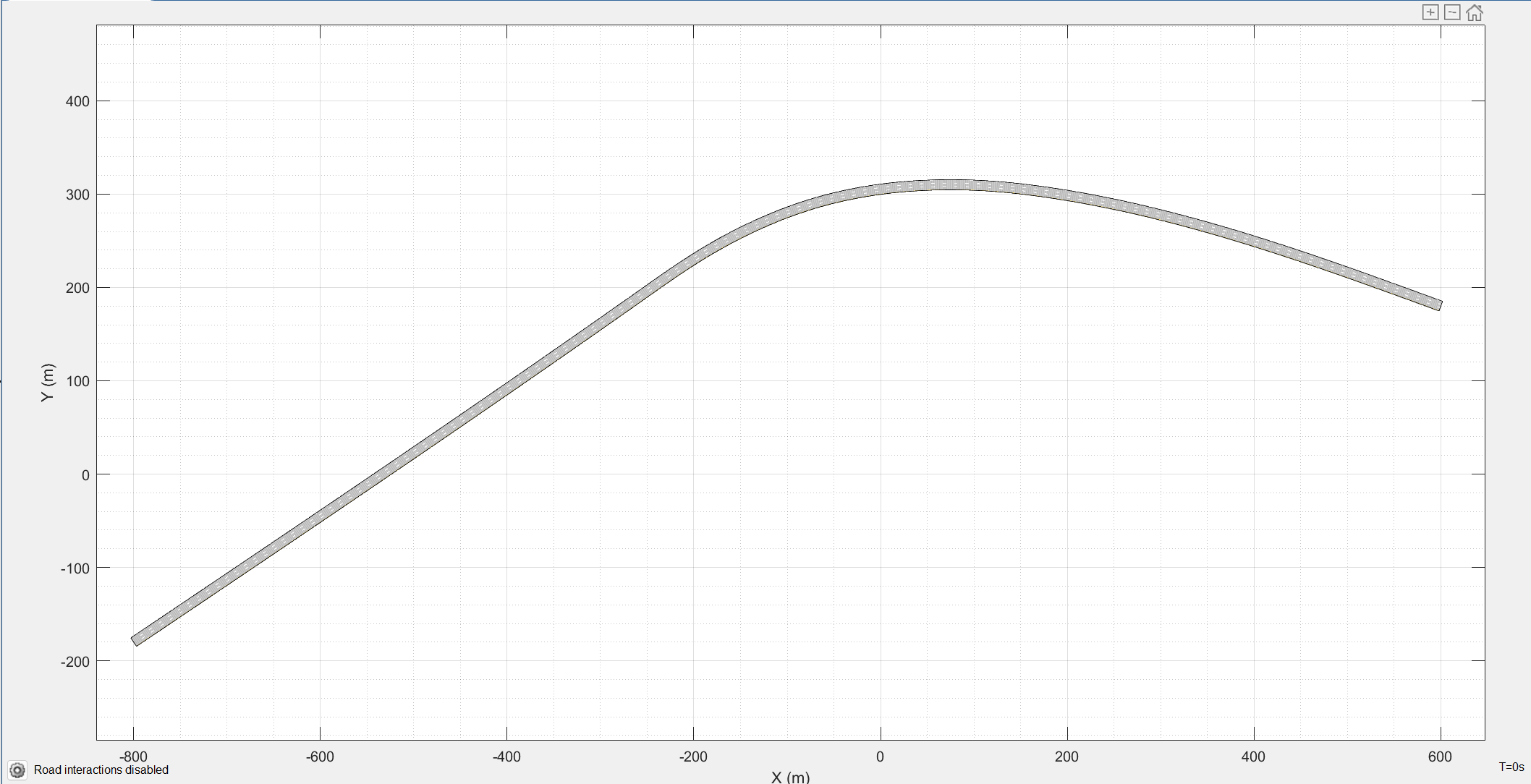}
    \label{fig:A4Bergamoreplica}    
    }
    \caption{A4 Bergamo satellite view and corresponding replica.}
    \label{fig:A4}
\end{figure*}

\begin{table*}[t]
    \caption{Description of the requirements for our running example.}
    \label{tab:requirements}
    \centering
    \begin{tabular}{l l p{15cm}}
    \toprule
    \textbf{Req.}   &     \textbf{Comp.} & \textbf{Requirement Description} \\
    \midrule
     R1    &  ACC & The longitudinal distance between the ego vehicle and the preceding vehicle shall be higher than a safety threshold assigned by the safety engineers (e.g., $20$m in our experiments).\\
     R2    &   LKA & The lateral deviation between the position of the vehicle and the center of the roadway should be at most one meter.\\
    \bottomrule
    \end{tabular}
\end{table*}

\begin{figure*}
\centering
\subfigure[Test Case T1.]{
    \includegraphics[width=.9\columnwidth]{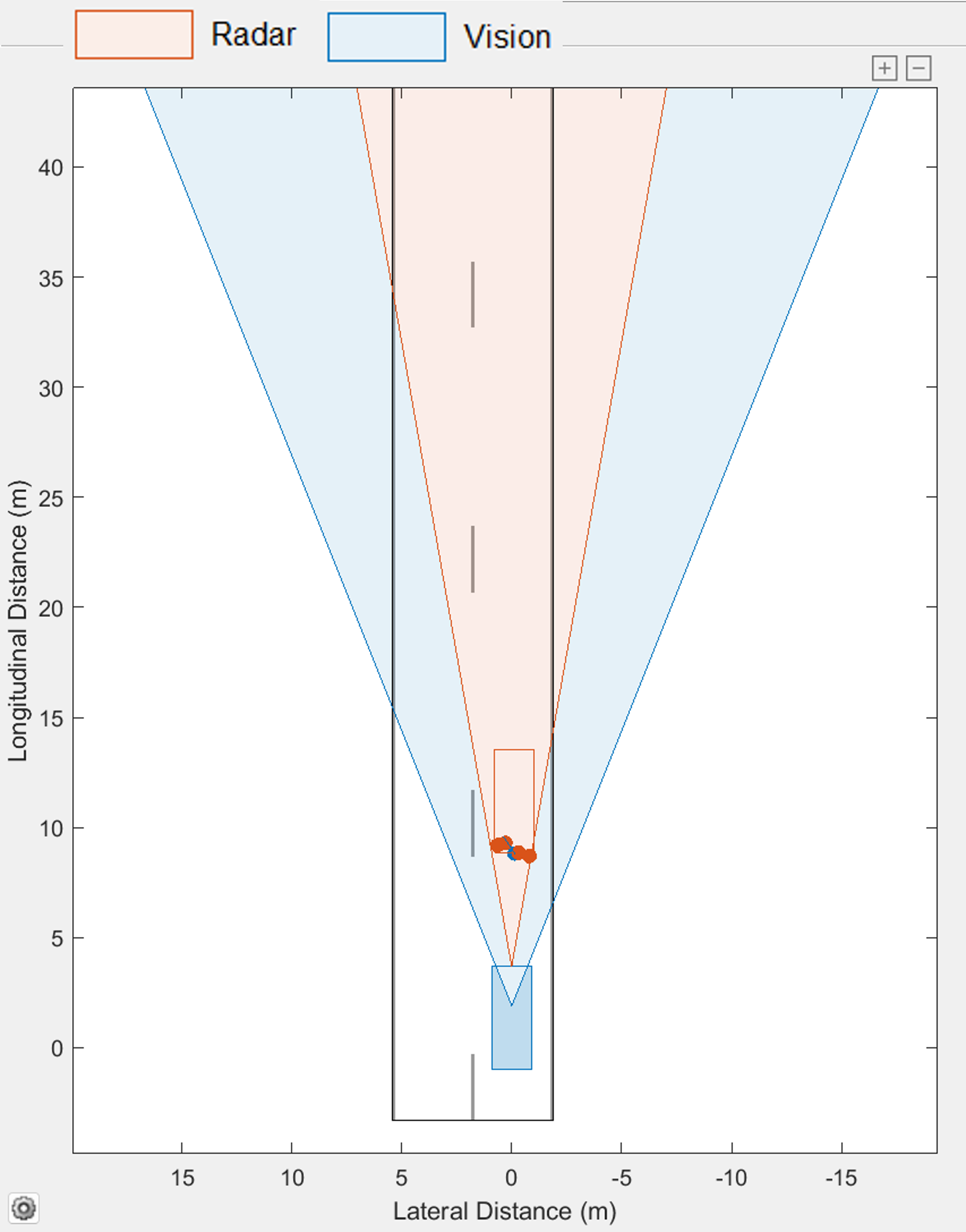}
\label{fig:testcase1}}
\quad
\subfigure[Test Case T2.]{
    \includegraphics[width=.9\columnwidth]{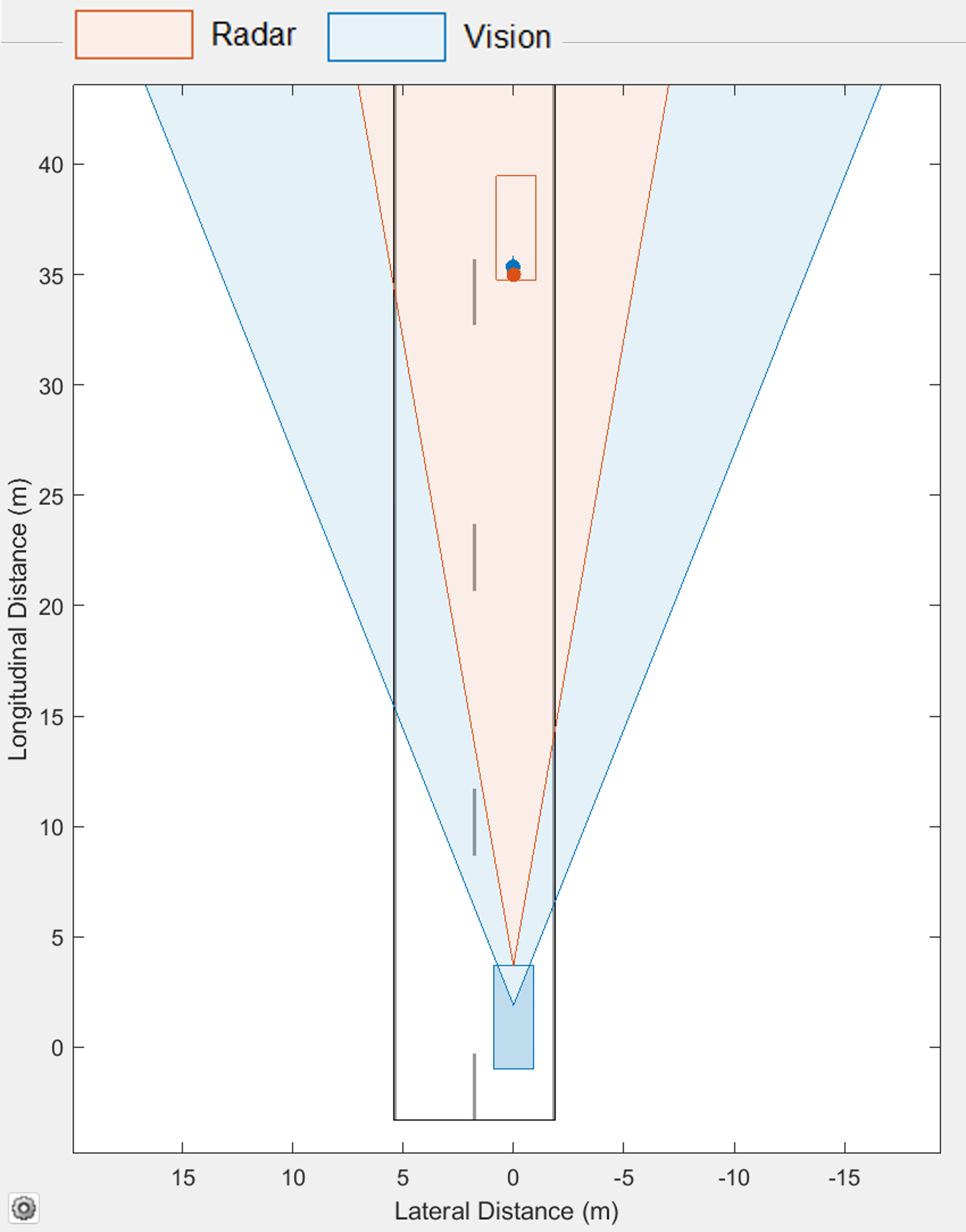}
    \label{fig:testcase2}    
    }
    \caption{Two test cases for our ADAS system.}
    \label{fig:testCases}
\end{figure*}

The \emph{Vehicle and Environment} component in \Cref{fig:SimulinkModel} initializes and executes the simulation. 
It considers both the \emph{Acceleration} and \emph{Steering Angle} of the ego vehicle, as defined by the \emph{Line Following Controller}, and the driving scenario, including the vehicle configurations and the road environment in which the simulation takes place. 
During execution, this block continuously provides the current state of the vehicle within the scenario, such as its position and dynamic behavior.
This component provides as outputs all signals measured by the simulated vehicle, including perception information (\emph{Vision}--from the cameras and \emph{Radar}), the measured \emph{Longitudinal Velocity}, the \emph{Lane Sensor}, as well as the position of all \emph{Actors} involved in the simulation.

\Cref{fig:Bird's-Eye Scope} shows an example of the 2D visualization generated during the simulation by the \emph{Vehicle and Environment} component. 
The visualization is divided into two views. 
\Cref{fig:Bird's-Eye Scope Vehicle coordinates view} reports the vehicle-coordinate view, where the ego vehicle is used as the reference frame. 
The figure shows the detected objects (\emph{Vision} and \emph{Radar}), lane boundaries, lane detection (\emph{Lane Sensor}), and the \emph{Tracks} associated with the surrounding vehicles.
The vehicle selected as the Most Important Object (\emph{MIO}) is highlighted with a yellow square. 
\Cref{fig:Bird's-Eye Scope World coordinates view}  reports the world-coordinate view, where the same scenario is represented in the global reference frame. This view shows the position of the ego vehicle and the surrounding vehicles along the road. 
Overall, this visualization allows engineers to inspect the current simulation state, verify the objects detected by the sensors, and identify the vehicle considered the most relevant by the controller during the execution.


The \emph{Collision Detection} component monitors the occurrence of collisions during the simulation, based on other \emph{Actors} in the same simulation and on the information provided by the \emph{Lane Sensor}. If a collision is detected, this block terminates the simulation early and marks the corresponding test case as failed. This allows the model to automatically identify failure-revealing simulations without requiring the execution to continue until the predefined simulation end time.

\medskip
\textbf{Requirements.}
For illustration purpose, we have used some simple requirements, given in \Cref{tab:requirements}, to be satisfied by our ADAS.
The requirement associated with ACC concerns the longitudinal distance from the preceding vehicle. 
Throughout the simulation, the ego vehicle must maintain a safe distance from the preceding vehicle. 
In particular, we require the \textit{distance} signal to never fall below 20 meters. 
If, at any simulation step, this value becomes lower than 20 meters, the corresponding test case is considered a failure-revealing test case.
This ensures that the ACC controller does not allow the vehicle to get too close to the preceding vehicle, which could lead to unsafe driving conditions and potential collisions. 
The requirement associated with the LKA concerns the lateral deviation of the ego vehicle with respect to its lane. 
The \textit{lateral deviation} signal measures the position of the vehicle inside the lane. 
In our model, the deviation between the position of the vehicle and the center of the roadway should be at most one meter. 
If the lateral deviation exceeds one meter at any point during the simulation, the corresponding test case is considered a failure-revealing test case. 
This requirement ensures that the LKA controller keeps the vehicle within an acceptable lateral displacement.

\begin{table*}[t]
    \centering
    \caption{Characteristics of the vehicles modeled by our SPL.}
    \label{tab:splvehicles}
    \begin{tabular}{clcccccc}
         \toprule
         \textbf{ID} & \textbf{Vehicle} & \textbf{M $[kg]$} & \textbf{YMI $[mNs^2]$} & \textbf{$\mathbf{D_f}$ $[m]$} & \textbf{$\mathbf{D_r}$ $[m]$} & $\mathbf{a_{max}}$$[m/s^{2}]$ & $\mathbf{a_{min}}$$[m/s^{2}]$\\ \midrule
         V1 & City car 1 & 1181 & 1919 & 1.275 & 1.275 & 2.89 & -11.63 \\ 
         V2 & Compact SUV & 1270 & 2890 & 1.275 & 1.275 & 2.67 & -10.57 \\
         V3 & City Car 2 & 1055 & 1393 & 1.150 & 1.150 & 2.00 & -9.37 \\
         V4 & Sedan 1 & 1422 & 2875 & 1.420 & 1.420 & 3.56 & -6.00 \\
         V5 & Pickup Truck & 2062 & 7859 & 1.630 & 1.630 & 3.52 & -9.30 \\
         V6 & Sport Car & 1772 & 3499 & 1.400 & 1.400 & 7.94 & -9.80 \\
         V7 & Sedan 2 & 1580 & 3344 & 1.400 & 1.400 & 3.91 & -11.30 \\ \bottomrule
    \end{tabular}
\end{table*}

\subsection{The Case Study Software Product Line}
\label{sec:SPL}
We consider a software product line (SPL) obtained by considering seven different vehicles, 22 driving scenarios, and two software components.

We considered the seven vehicles from \Cref{tab:splvehicles}.
These vehicles are from four different manufacturers, 
selected to represent a range of market segments, including city cars, sedans, compact SUVs, off-road vehicles, and sports cars.
The considered vehicles differ in terms of mass (\emph{M}), yaw moment of inertia (\emph{YMI}), longitudinal distance from the center of gravity and the front tires ($D_f$) and from the rear tires ($D_r$), and the maximum ($a_{max}$) and minimum ($a_{min}$) acceleration.

We considered the $22$ realistic \textbf{driving scenarios}.
A \emph{driving scenario} specifies the road geometry, traffic participants, and environmental conditions under which the ADAS is simulated. 
Each scenario defines the operating conditions used during testing.
These scenarios challenge the two ADAS functionalities.
Among the scenarios, seven (S1-7) come from the original \matlab/\Simulink, three (S8-10) were created to stress specifically one aspect of the ADAS, and $12$ (S11-22) are exact replicas of excerpt coming from existing roads around the world (i.e., ``Autostrada Trafori'', ``Overseas Highway '', ``Pacific Coast Highway'', ``Sea Sky Hwy'', ``Trans Canada Highway'', ``A8 Stuttgart Munich'', ``I\_81 Highway'', ``A4 Bergamo'', ``A15 LaSpezia Parma'', ``Highway double target'', ``Queen Elizabeth Way Oakville'', ``Las Vegas Freeway''). 
Real-world scenarios are primarily high-speed road environments and are mapped through satellite map data via the \matlab Driving Scenario Designer~\cite{DrivingScenario}.
For example, \Cref{fig:A4} shows an excerpt of the A4 high-speed road from Google Maps (\Cref{fig:A4BergamoSatellite}) and the corresponding replica obtained with \matlab Driving Scenario Designer (\Cref{fig:A4Bergamoreplica}). 

The two software components are related to two ADAS functionalities: ACC which maintains a target speed and a safe distance from the preceding vehicle, and LKA which helps keep the vehicle within lane boundaries. 
Both systems are safety-critical, as incorrect behavior may lead to collisions or loss of vehicle control.

By combining these three dimensions, we obtained a total of $308$ (7$\cdot$22$\cdot$2) distinct products to be tested, i.e., \textbf{test cases}.
We consider a \emph{test case} to be the execution of a given software configuration (i.e., a vehicle configuration and software component version) under a specific driving scenario, together with the associated test oracle that checks the system requirements using \Simulink Test Assessment blocks. A test case, therefore, corresponds to one simulation run.
Among those, we considered $224$ baseline test cases within our \textbf{test suite} since, for some scenarios (i.e., those with straight roads coming from the documentation), the LKA is not tested as it is not extensively exercised.
As a consequence, such scenarios are not used when tests target the LKA component.
Therefore, our \emph{test suite} is the collection of all baseline test cases considered in this work.
Among these test cases, $133$ refer to the ACC software component, and $91$ to the LKA.
Executing the full test suite of $224$ test cases requires approximately $59$ hours of simulation time. 
This cost makes exhaustive regression testing impractical in OTA contexts, where updates must be validated and deployed rapidly.

Each product is evaluated against the set of requirements reported in \Cref{tab:requirements}, which are specified as \Simulink Test Assessment blocks~\cite{TestSequences}.
It is worth noting that all  $224$ baseline test cases satisfy the specified requirements: no collision occurs and the lateral deviation remains within the accepted range.


\subsection{The Case Study OTA Updates}
\label{sec:ProblemDefinitionOTAs}
We considered five OTA updates affecting the behavior of the two ADAS software components:
two targeting ACC and three targeting LKA. 
\Cref{tab:OTAs} describes the OTA updates considered in this work.
Therefore, there are 11 possible software configurations in addition to the original software configuration (3 configurations of the ACC $\times$ 4 configurations of the LKA minus the original configuration). 
Although each update modifies only a subset of the system, its effects may propagate across multiple configurations, making it non-trivial to determine which test cases are necessary to ensure safety.

\begin{table*}[t]
    \caption{Description of the five OTA updates for our running example. }
    \label{tab:OTAs}
    \centering
    \begin{tabular}{l | l | p{.78\textwidth}}
    \toprule
    \textbf{ID}   &     \textbf{Component} & \textbf{Description} \\
    \midrule
      OTA1 & ACC & Limits the ACC controller to use only $10\%$ of its total braking force.\\
      OTA2 & ACC & Reduces the distance the ACC maintains between the ego vehicle and the preceding vehicle from $20$m to $10$m and the maximum deceleration the ACC can apply. \\ 
      OTA3 & LKA &  Limits the LKA controller to use only the 10\% of the total steering capability.\\
      OTA4 & LKA &  Reduces the controller prediction horizon from 30 to 7 (i.e., we pass from predicting next 30 actions to be applied to just 7). \\ 
      OTA5 & LKA &  Applies a gain of 0.1 to the preview curvature signal (the centerline of lane curvature ahead of the ego vehicle). The controller uses previewed information for calculating the ego vehicle steering angle, which influences the controller performance. \\
    \bottomrule\end{tabular}
\end{table*}

When an OTA update is introduced in a CPS,  some test cases become failure-revealing, since the modified controller is no longer able to satisfy the requirements specified in the test assessment. 
Therefore, regression testing is required to ensure that the updated system does not violate safety or functional requirements. 
Re-executing the full test suite ($224$ test cases --- $59$ hours of simulation time) for each software configuration may not be possible.
Given the large size of the test suite and the time constraints typical of OTA deployment pipelines, executing all test cases in arbitrary order may delay the detection of critical failures. 
Therefore, an effective prioritization strategy is needed to maximize the likelihood of revealing failures early during test execution. Therefore, our work considers the following problem.

\begin{Answer}[Problem~1] How to effectively support safety analysis of OTA updates for complex software product lines by \emph{prioritizing} the test cases from the regression test suite?
\end{Answer}

Despite the fact that an effective prioritization strategy may help maximize the likelihood of revealing failures early during test execution, re-executing the full test suite for each software configuration may not be possible. 
In these scenarios, engineers must identify the subset of test cases to be re-executed after each OTA update.
As a consequence, we consider the following additional problem.

\begin{Answer}[Problem~2] How to effectively support safety analysis of OTA updates for complex software product lines by \emph{selecting} a subset of the test cases from the regression test suite?
\end{Answer}






%% file: sections/testselectionandprior.tex
\section{Criticality-Based Support for Safety Analysis of OTA Updates} \label{sec:method}
This section presents \crisp, our criticality-based test case prioritization  (\Cref{sec:prioritization}) and selection (\Cref{sec:selection}) approach. 
The key idea behind \crisp is to treat criticality values computed during baseline verification as reusable artifacts that guide regression testing after subsequent OTA updates.


\subsection{Test Case Prioritization} \label{sec:prioritization}
The test case prioritization technique implemented by \crisp is based on Hypothesis~1 introduced in \Cref{hyp1}.
Intuitively, this hypothesis suggests that more critical baseline test cases are more likely to expose failures after an OTA update. 
For example, in our ADAS case study, executing the test case in \Cref{fig:testcase1} before \Cref{fig:testcase2} is advantageous, given it is more critical.


Our approach to validating this hypothesis consists of the following steps.

\textit{Step~1}. When a baseline test case $t$ is executed (e.g., before the OTA update), \crisp computes a fitness value $f(t) \in \mathbb{R}$ where:
\begin{enumerate}
    \item a negative value indicates that at least one requirement has been violated;
    \item a positive value indicates that all requirements are satisfied;
    \item a higher positive value means the system is further from violating its nearest-to-violation requirement; and 
    \item a lower negative value means the system is further from satisfying its furthest-from-satisfaction requirement.
\end{enumerate}
Test cases with smaller fitness values are considered more critical.
For example, the two test cases from \Cref{fig:testCases} are associated with the fitness values 5 and 30.

\textit{Step~2}. Test cases are prioritized from most to least critical by sorting them in ascending order of their fitness values.
%
We hypothesize that this ordering ensures that test cases most likely to expose failures are executed earlier, and 
our evaluation will empirically validate the effectiveness of our solution. 
For example, the test case from \Cref{fig:testcase1} is executed before the test case from \Cref{fig:testcase2} since it has a lower fitness value. As illustrated, the ego vehicle in \Cref{fig:testcase1} is in a more critical situation than the one in \Cref{fig:testcase2}. While this difference is visually apparent from the simulation snapshots, it is quantitatively captured by the fitness value computed during the simulation. In this case, the vehicle is close to an unsafe condition and close to violating the specified requirements, resulting in a low fitness value for the corresponding simulation. 
In the second case, instead, the ego vehicle is in a safer configuration and remains farther from the requirement violation. This results in a higher fitness value, indicating that the simulation is less critical.

\smallskip

\noindent\textbf{Implementation.}
To compute \emph{fitness} values, we leverage \Hecate~\cite{10374027}, a research toolbox supporting SBST for \Simulink models.
This toolbox generates test cases and evaluates their criticality based on a fitness function from \Simulink Test Assessments~\cite{TestAssessment} blocks. 
In this work, we exploit the fitness generation component offered by \Hecate.
\Hecate has been successfully applied to support the development of a cruise controller for an industrial simulator~\cite{10.1145/3611643.3613894}, a drone controller~\cite{Bisceglia2026-pe}, and the software controller of an e-bike~\cite{10.1007/978-3-032-24839-8_8}.  

Although \Hecate can be used to generate test cases, in this work we only partially exploit its functionalities.
More specifically, we only consider the formalization of requirements via Test Assessments and we use \Hecate to compute the \emph{fitness} (or \emph{robustness}).
For each test case simulation, \Hecate evaluates the deviation between observed signals and the thresholds specified in the Test Assessment, returning the minimum distance to a violation observed during the execution. 
This value is used as the \emph{fitness} $f(t)$.
Therefore, we use \Hecate to compute the fitness function from the \Simulink Test Assessment Blocks used to specify the requirements (see \Cref{sec:SPL}).
We remark that fitness values are system-dependent and are computed once on the baseline system, prior to the deployment of OTA updates.
This design choice enables the computed fitness values to be reused across subsequent updates, eliminating the need for repeated computation and improving the scalability of \crisp.
\smallskip

While test case prioritization aims to detect failures earlier, executing the entire test suite $T$ may still be infeasible under strict time constraints typical of an OTA. 
Therefore, it is fundamental to develop effective test case selection solutions, such as the one we proposed in the next section.

\subsection{Test Case Selection} \label{sec:selection}
\crisp introduces a test case selection strategy that reduces the number of simulations while preserving the ability to detect failures effectively.
Our approach exploits similarities in system behavior across different vehicle configurations within the same driving scenario to select the test cases to be re-executed after an OTA. 
Intuitively, the test case selection technique implemented by \crisp is based on the 
Hypothesis~2 introduced in \Cref{hyp1}.

Given a set of test cases ($T_s$) for scenario ($s$), if the criticality difference across various test cases (i.e., $\Delta(s) = \max_{t \in T_s} f(t) - \min_{t \in T_s} f(t)$) is lower than a threshold ($\theta$), the test cases behave similarly. Therefore, only the most critical test case is executed.
More precisely, given a threshold parameter ($\theta$), \crisp operates as follows. If the criticality difference falls below the threshold (i.e., $\Delta(s) \leq \theta$), the behavior of different vehicle configurations is considered sufficiently similar, in which case, only the most critical test case is selected, i.e., the test case with the smallest fitness value. 
Otherwise, the behavioral differences across configurations are significant, and all test cases for that configuration are selected for execution.

For example, in the \textit{Pacific\_Coast\_Highway} scenario (S13), the difference in criticality among the seven considered vehicles is $\Delta(s) = 0.94$. 
This value is below the selected threshold ($\theta=5$), so the vehicles are considered sufficiently similar. 
Therefore, instead of simulating all seven vehicles, we only simulate the most critical one (i.e., V5 in \Cref{tab:splvehicles}). 
The test outcome for this vehicle is then considered representative of the remaining six vehicles in the same scenario. 

\smallskip
\noindent\textbf{Implementation.} Fitness values are computed by \Hecate, which returns quantitative measures of distance to violation of the requirements specified within Test Assessments.
The selection is performed by considering the fitness values from the baseline test cases and subsequently applied to guide test execution after OTA updates.

\smallskip
We posit that similarity in baseline behavior is likely to indicate similarity in failure-revealing potential across configurations, particularly in the absence of significant structural changes introduced by the update.
We  analyze the effectiveness of this selection strategy in \Cref{sec:evaluation}.

%% file: sections/evaluation.tex
\section{Evaluation} \label{sec:evaluation}
We considered the following research questions to evaluate our approach:
\begin{itemize}
	\item \textbf{Efficiency (RQ1)}: How effective is our test case prioritization approach, introduced in \Cref{sec:prioritization},  in finding failure-revealing test cases? (\Cref{sec:rq1})
	\item \textbf{Effectiveness (RQ2)}: How effective is our test case selection approach, introduced in \Cref{sec:selection} , in extracting failure-revealing test cases? (\Cref{sec:rq2})
\end{itemize}

\subsection{RQ1: Effectiveness of Test Case Prioritization}\label{sec:rq1}

\textbf{Methodology}. Given the example from \Cref{sec:background}, we used two baseline techniques.
\begin{itemize}
    \item \emph{Random Ordering}. It executes all the test cases in random order. 
\item \emph{Anti-patterns based (\apdisc)}. It prioritizes test cases that produce short and abrupt peaks in the output signals during execution (i.e., the discontinuity antipattern~\cite{Arrieta2026}).
The assumption is that such discontinuities are more likely to reveal anomalous or failure-prone system behaviors. 
\end{itemize}
We chose \emph{Random Ordering} to represent the absence of any informed test execution strategy, and \emph{\apdisc} to represent the state-of-the-art technique.
Indeed, \apdisc was identified as the best test prioritization technique for CPSs modeled in \Simulink in a recent work~\cite{Arrieta2026}.

We considered each of the five OTA updates. 
We re-executed our $224$ test cases using our prioritization techniques and the two baseline techniques.
The five OTA updates were introduced individually, without overlapping modifications: Two updates targeted the ACC component (OTA1 and OTA2), while the remaining three affected the LKA component (OTA3, OTA4, and OTA5).
We repeated the experiments related to the Random Order baseline five times to account for the randomness of this prioritization strategy.

To compare our solutions, we considered two metrics.
\begin{itemize}
    \item The number of test cases executed by the techniques before identifying the first failure-revealing test case. We refer to this metric as the Time to First Failure (TTFF).\footnote{We use the number of test cases as a proxy for the TTFF since, for our model, all the test cases take approximately the same time to be executed.}
This metric measures the effectiveness of the different solutions in identifying \emph{the first} failures.
The intuition is that since engineers are asked to deploy an OTA that offers sufficient \emph{safety} guarantees, no failure-revealing test cases must be present.
Therefore, after identifying one failure-revealing test case, engineers should fix their model.
For each technique, we compute the average number across our OTAs of test cases executed for identifying the first failure-revealing test case.
The lower the value, the more effective the solution.
\item The Average Percentage of Faults Detected (APFD), a largely used metric~\cite{988497,doi:https://doi.org/10.1002/9781119896838.ch7,10429997}  which measures \emph{when} the different failure-revealing test cases are revealed. 
The APFD metric assumes values between 0 and 1; higher APFD values indicate that failure-revealing test cases are detected earlier during test execution.
The higher the value, the more effective is the prioritization strategy.
\end{itemize}



To verify whether our results are statistically significant, we applied the two-sample Kolmogorov–Smirnov (K–S) test~\cite{doi:https://doi.org/10.1002/9781118445112.stat06558}.
It is a non-parametric test that compares the entire distributions of two samples without assuming normality, making it suitable for the distributions of failure-revealing test-case indexes produced by different prioritization strategies. 
To complement the significance analysis, we computed the Vargha–Delaney $\hat{A}_{12}$ effect size~\cite{Neumann2015}, a non-parametric measure that quantifies the probability that one strategy detects failures earlier than another, thereby indicating the magnitude and direction of the observed differences.

\addtolength{\tabcolsep}{-0.1em}
\begin{table}[t]
	\centering
	\caption{TTFF metric among different approaches.}
	\label{tab:FirstFailure}
	\begin{tabular}{l | r r r r r | r}
		\toprule
		 & $\mathbf{OTA1}$ & $\mathbf{OTA2}$ & $\mathbf{OTA3}$ & $\mathbf{OTA4}$ &  $\mathbf{OTA5}$ & $\mathbf{Avg}$ \\ \midrule
		\crisp & 3 & 9 & 1 & 1 & 1 & 3.0 \\
		\apdisc & 2 & 3 & 1 & 1 & 1 & 1.6 \\
		Random & 3.2 & 2.8 & 4.8 & 2.6 & 13.2 & 5.32 \\
        \bottomrule
	\end{tabular}
\end{table}

To mitigate possible biases introduced by a single random permutation, the random strategy was repeated five times using different execution orders, and the reported results correspond to the average across the five runs.

\textbf{Results}. Our results show that $46$ and $23$ out of $133$ test cases are failure-revealing for the OTA updates related to ACC (OTA1 and OTA2, respectively) and that $33$, $25$, and $7$ out of $91$ test cases are failure-revealing for the OTA updates related to LKA (OTA3, OTA4, and OTA5, respectively).

\Cref{tab:FirstFailure} reports the number of test cases the different approaches need to execute before finding the first failure-revealing test case.
The results show that \crisp and \apdisc outperform Random Ordering for four out of five OTAs.
Unlike the other four OTAs, the change introduced by OTA2 is quite disruptive, as it directly changes the distance the ACC maintains between the ego vehicle and the preceding vehicle from $20$m to $10$m and the maximum deceleration the ACC can apply (see \Cref{tab:OTAs}). 
This change directly impacts requirement satisfaction: the longitudinal distance between the ego vehicle and the preceding vehicle shall be greater than $ 10$ m (see R1 in \Cref{tab:requirements}). 
Therefore, since this change has a disruptive impact on the ACC's functionality, \emph{Random Ordering} yielded the best results, given the driving capabilities of the test cases' criticality and the anti-pattern-based approach.
For OTA1, \apdisc outperforms \crisp. 
Unlike \crisp, which uses the distance between the ego vehicle and the preceding vehicle (derived from the requirements from \Cref{tab:requirements}) to compute the fitness function, \apdisc uses short and abrupt peaks in the output acceleration signals (the distance between the ego vehicle and the preceding vehicle) to prioritize the test cases. 
Therefore, \apdisc orders the test cases by focusing on abrupt peaks in the relative acceleration between the vehicles.
This metric offers a better driver for prioritizing test cases related to OTA1, which limits the ACC controller to use only $10\%$ of its total braking force.
For OTA3, OTA4, and OTA5, \crisp and \apdisc outperform \emph{Random Ordering} and offer the same effectiveness considering the number of test cases that need to be executed by the techniques before identifying the first failure-revealing test case.

\begin{figure}
	\centering	\includegraphics[width=0.9\linewidth,clip,trim={4.3cm 0cm 5cm 0cm}]{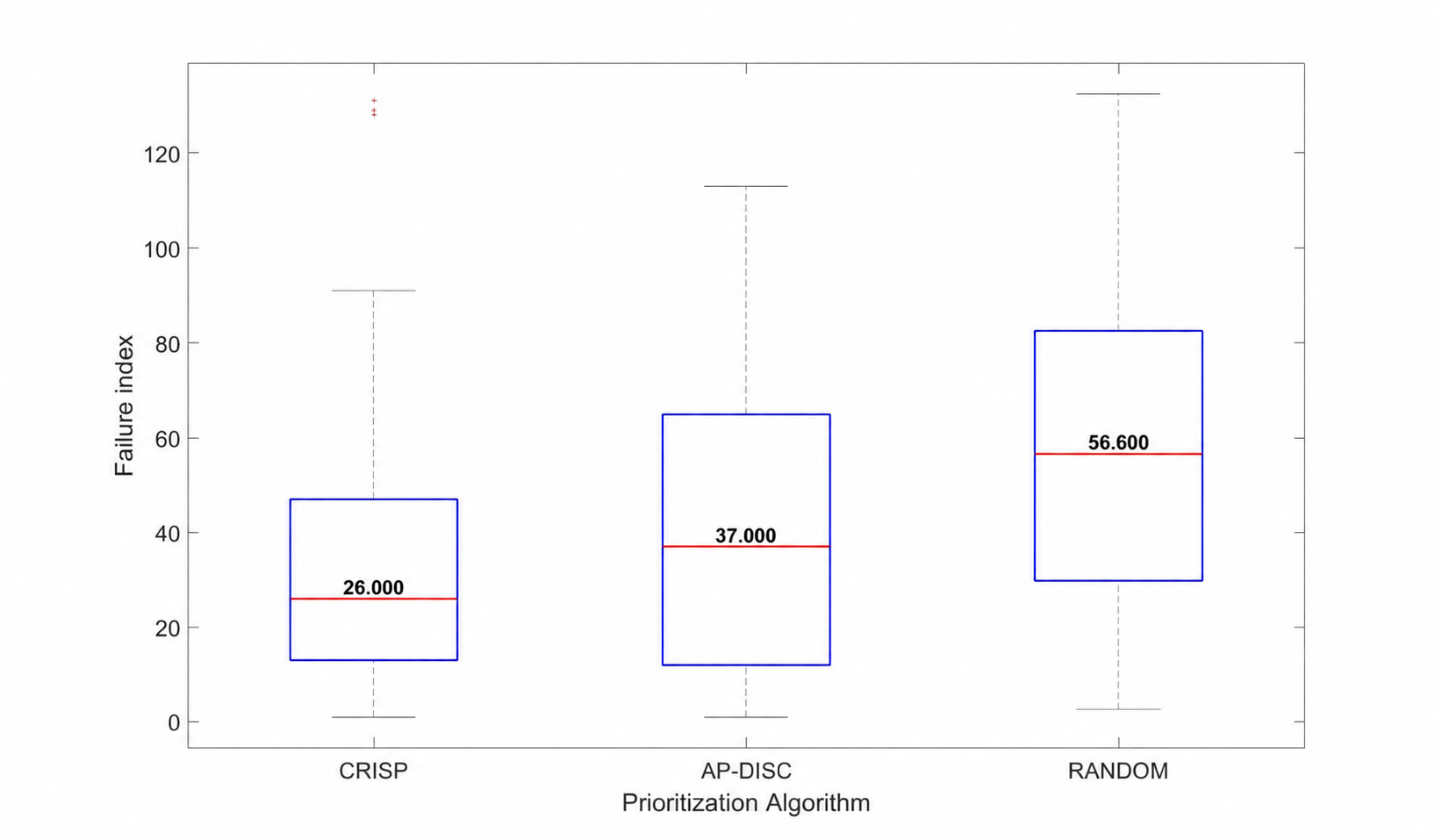}
	\caption{Average failure index per prioritization approach.}
	\label{fig:boxplot_RQ1}
\end{figure}

 \Cref{fig:boxplot_RQ1} shows the results from \crisp, \apdisc, and Random Ordering for the Average Percentage of Faults Detected (APFD) metric. 
Our results confirm that \crisp outperforms the baselines; the average lower failure index for \crisp, \apdisc, and Random Ordering is $26.0$ (min=$1$, max=$131$, std=$28.1296$), $37.0$ (min=$1$, max=$113$, std=$31.3665$), $56.6$ (min=$2.6$, max=$132.4$, std=$34.7526$), respectively.
The two-sample Kolmogorov-Smirnov and the Vargha and Delaney tests confirm that our results are statistically significant (p-value $4.7 \cdot 10^{-3}$ and average $\hat{A}_{12}=0.5549$).

\Cref{tab:APFD} reports the APFD values for each OTA update and test case prioritization technique. 
The results confirm that, on average, \crisp performs better than random and \apdisc.
Considering the OTAs separately, the results confirm that the criticality-based approach performs better than random and \apdisc in three cases out of five.
Surprisingly, for OTA1 and OTA2, unlike the number of test cases executed by the techniques before identifying the first failure-revealing test, for the APFD metric, \crisp outperforms \apdisc.
The opposite observation can be made for OTA4 and OTA5.

To summarize, our results show that \crisp and \apdisc generally outperform \emph{Random Ordering}, while offering complementary benefits depending on the considered metric. Since the APFD metric is considered a standard metric for comparing prioritization solutions, we consider the fitness function provided by \crisp for the rest of our experimentation.

\addtolength{\tabcolsep}{-0.1em}
\begin{table}[t]
	\centering
	\caption{APFD metric over the proposed robustness-based approach and the two baselines.}
	\label{tab:APFD}
    \footnotesize
	\begin{tabular}{l | ccccc | c}
		\toprule
		 & $\mathbf{OTA1}$ & $\mathbf{OTA2}$ & $\mathbf{OTA3}$ & $\mathbf{OTA4}$ &  $\mathbf{OTA5}$ & $\mathbf{Avg}$  \\ \midrule
		\crisp & \textbf{0.6734} & \textbf{0.7877} & \textbf{0.6828} & 0.6305 & 0.9333 & \textbf{0.7415} \\
		\apdisc & 0.6568 & 0.4637 & 0.6632 & \textbf{0.7026} & \textbf{0.9568} & 0.6886 \\
		Random & 0.4915 & 0.4569 & 0.4624 & 0.5256 & 0.5003 & 0.4873 \\ \bottomrule
	\end{tabular}
\end{table}

\begin{Answer}[RQ1 --- Answer]
	Our criticality-based prioritization approach is (on average) more effective than prioritizing test cases in a random order and the best approach from the literature (Anti-patterns based --- \apdisc) according to a recent survey on the topic.
\end{Answer}

\subsection{RQ2: Effectiveness of Test Case Selection}\label{sec:rq2}

\begin{figure*}
	\centering
	\subfigure[OTA 1.\label{fig:tradeOffACCOTA1}]{
    \includegraphics[width=0.65\columnwidth,clip,trim={3cm 0cm 3cm 0cm}]{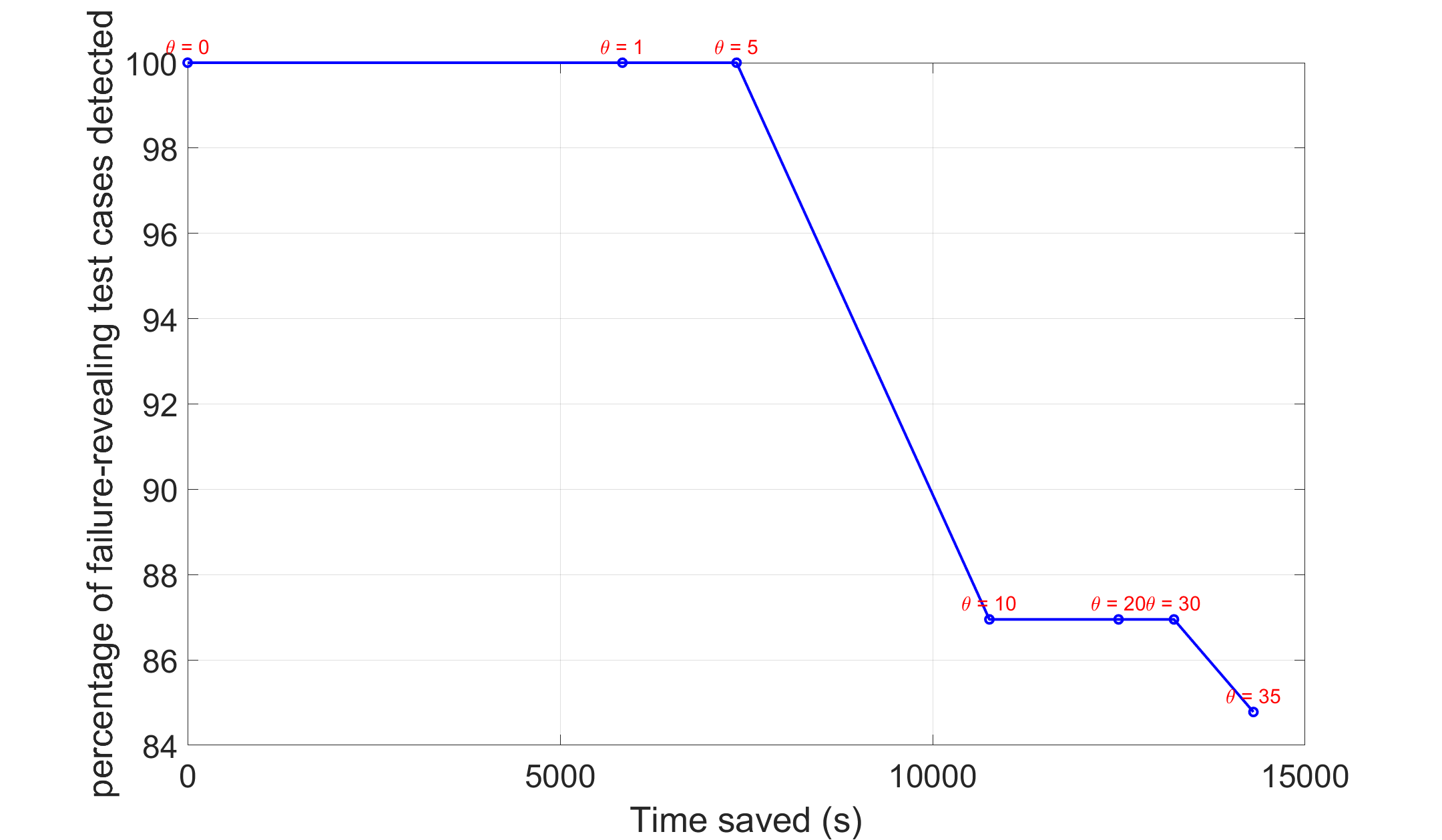}  
    }
    \subfigure[OTA 2.\label{fig:tradeOffACOTA2}]{\includegraphics[width=0.65\columnwidth,clip,trim={3cm 0cm 3cm 0cm}]{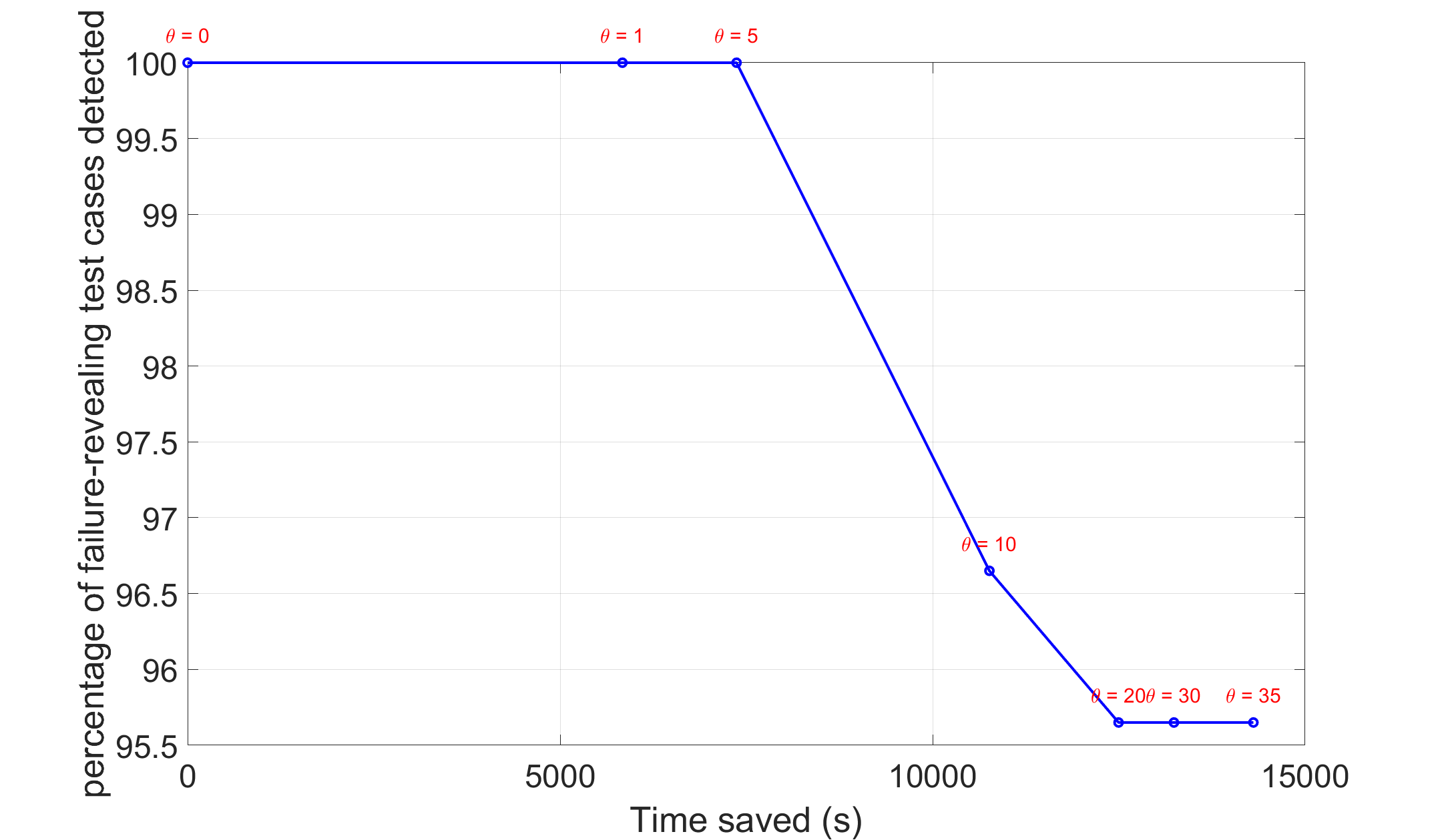}}
    \subfigure[OTA 3.\label{fig:tradeOffLKAOTA3}]{\includegraphics[width=0.65\columnwidth,clip,trim={3cm 0cm 3cm 0cm}]{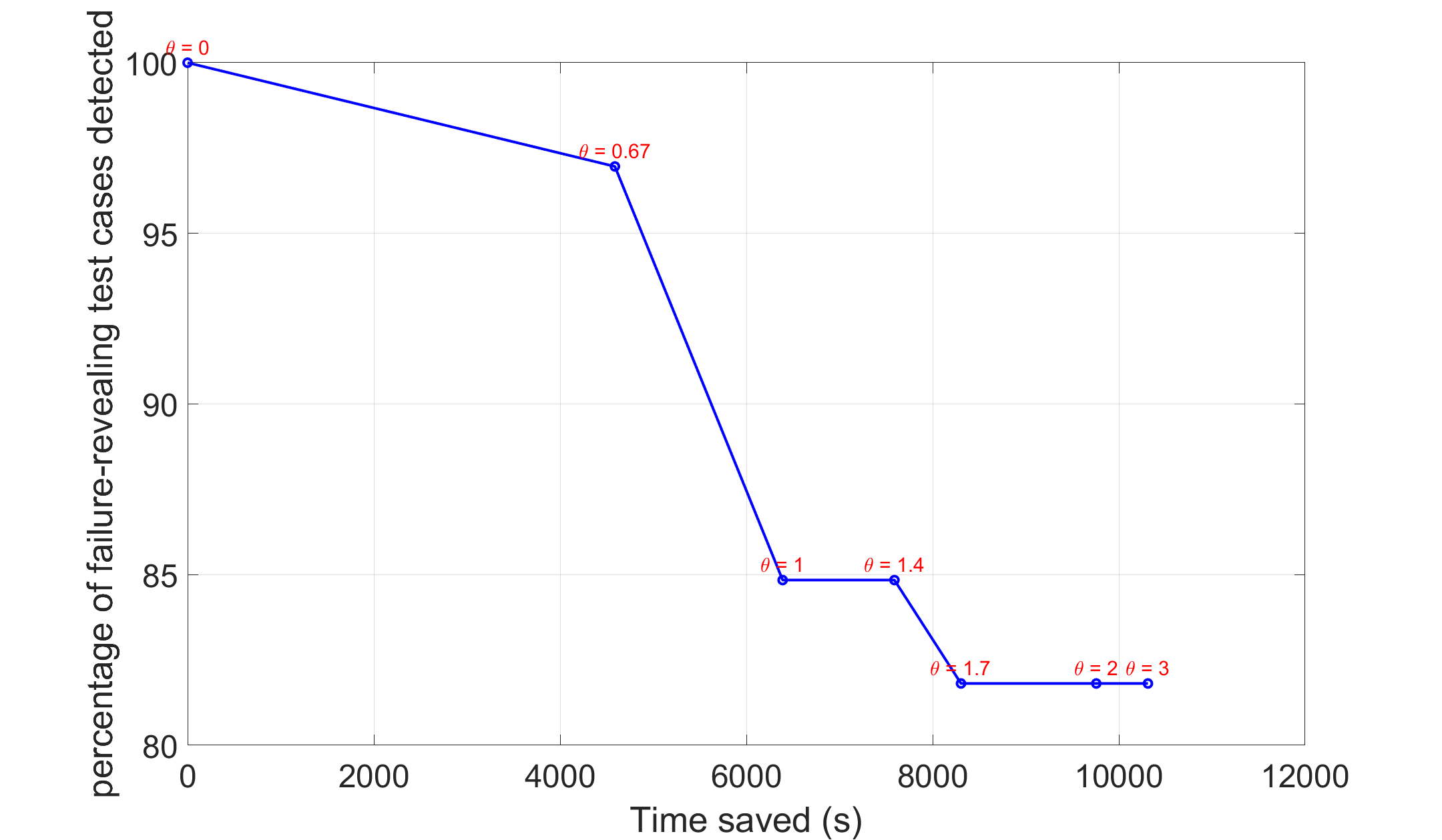}}
    \subfigure[OTA 4.\label{fig:tradeOffLKAOTA4}]{\includegraphics[width=0.65\columnwidth,clip,trim={3cm 0cm 3cm 0cm}]{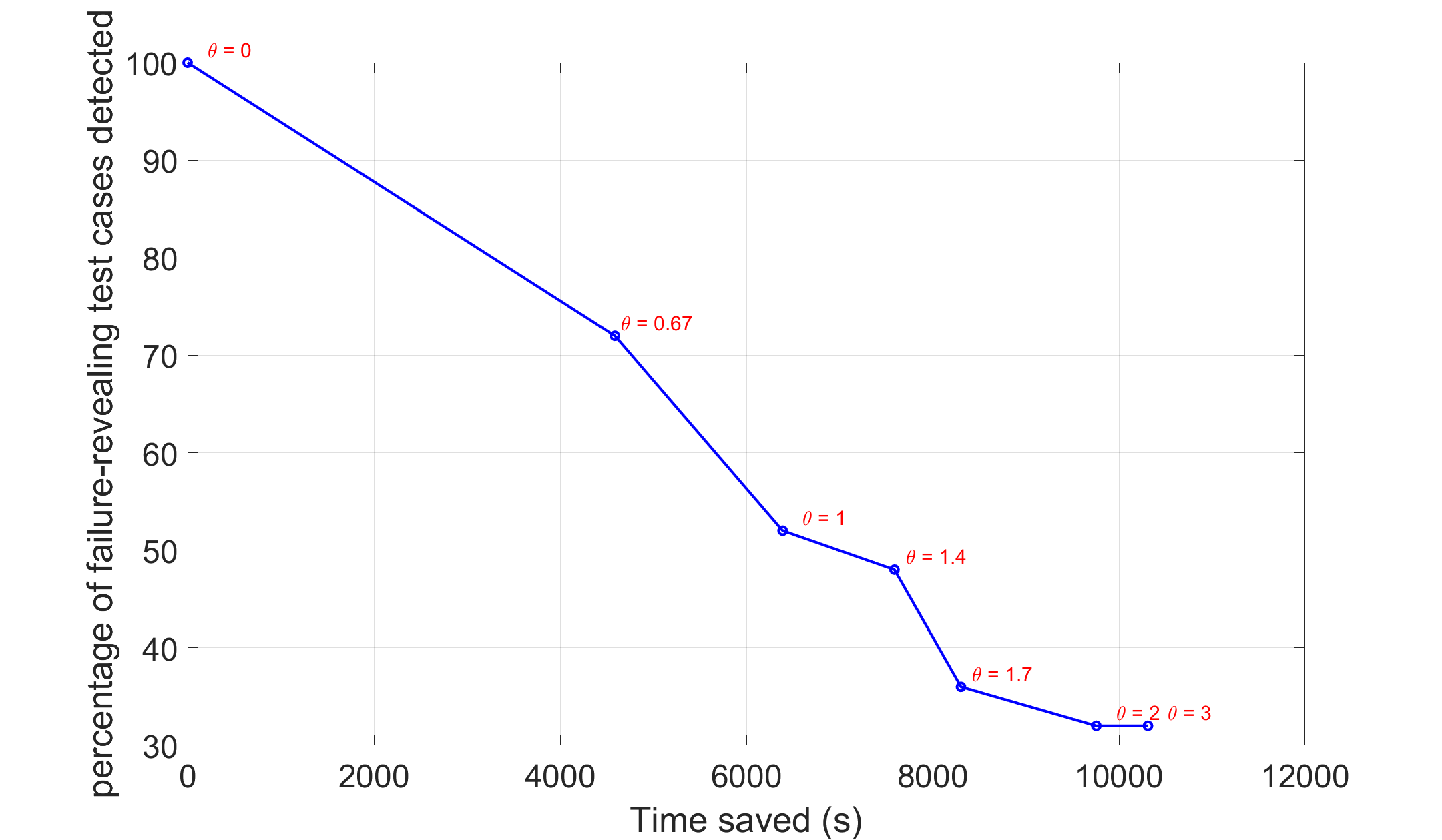}}
    \subfigure[OTA 5.\label{fig:tradeOffLKAOTA4}]{\includegraphics[width=0.65\columnwidth,clip,trim={2cm 0cm 3cm 0cm}]{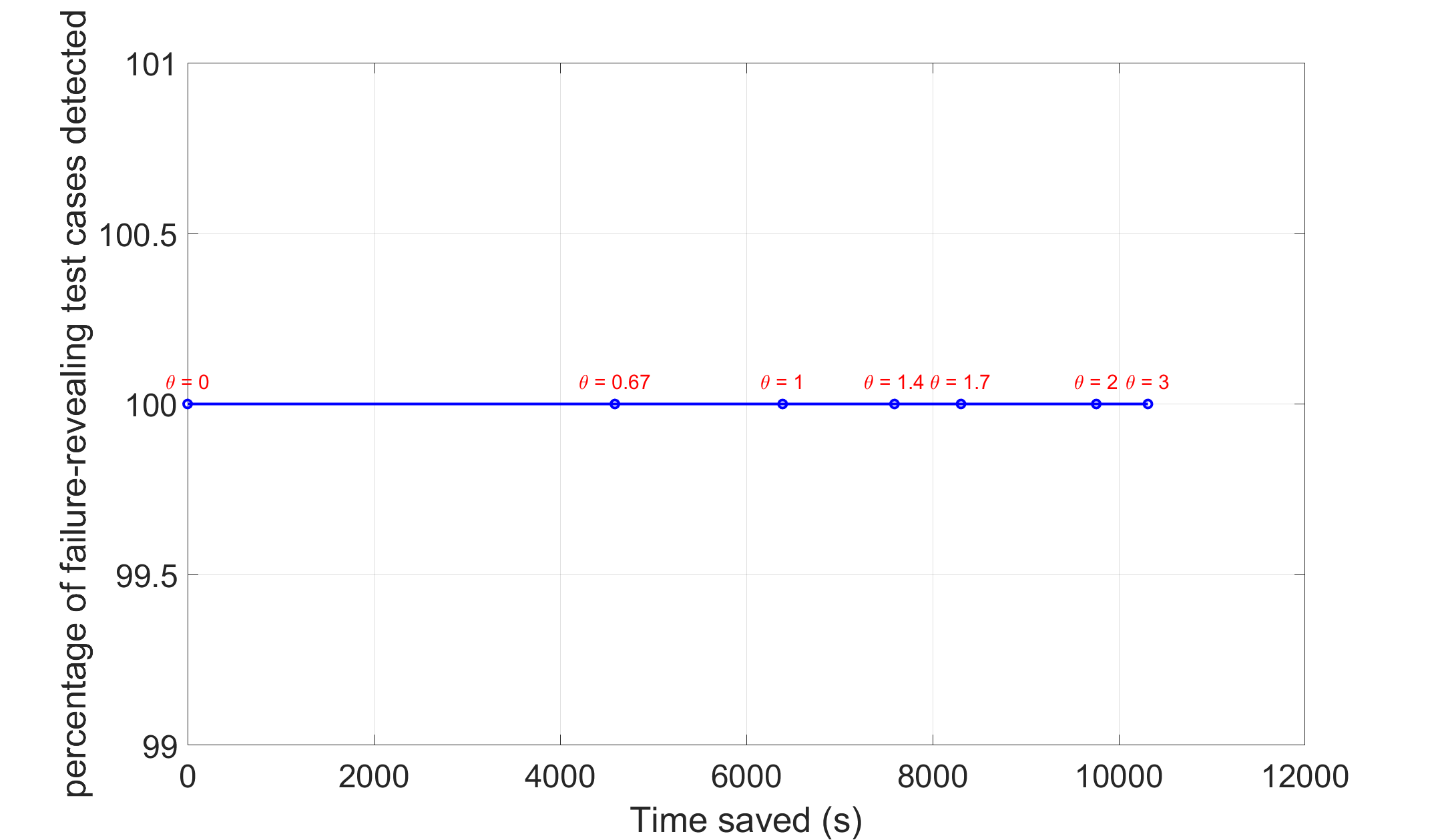}}
    \caption{Percentage of failures and time saving from \crisp for different values assigned to the threshold parameter ($\theta$).}
    \label{fig:ResultsRQ2TimeFailures}
\end{figure*}

In this section, we analyze the effectiveness of the test case selection procedure from \crisp. 
We are interested in assessing the capability of \crisp in identifying failure-revealing test cases and the time required.

\textbf{Methodology}.
We ran $35$ experiments, each performed for a given OTA (across our five OTAs) and threshold value ($\theta$).
We considered seven different threshold values since it affects the computation of similar behaviors, and we want to assess how this aspect affects the effectiveness of the test case selection procedure. 
We considered different values for the ACC-related OTAs and the LKA-related OTAs, because the range of the fitness values associated with the corresponding requirements (\Cref{tab:requirements}) differs.
For the ACC updates, the threshold varies between 0 and 35 (i.e., $0$, $1$, $5$, $10$, $20$, $30$ and $35$). 
For the LKA updates, the threshold varies between 0 and $3$ (i.e., $0$, $0.67$, $1$, $1.4$, $1.7$, $2$ and $3$). 
This allows the threshold analysis to be adapted to different fitness distributions observed for the two ADAS functionalities.

For each experiment, we measured (a)~the percentage of failure-revealing test cases detected (compared to the number of failure-revealing test cases identified in RQ1), and (b)~the time saved w.r.t. the execution of the entire test suite.
Test selection reduces the number of test cases by extracting one representative (the most critical) test case among several ones showing similar behaviors.
Furthermore, we consider this result to be a proxy for the remaining test cases (which are not executed).
Therefore, to compute the percentage of failure-revealing test cases, we assume that a failure-revealing test case is detected if (a)~it is executed (and confirmed as failure-revealing) or (b)~it is not executed and the corresponding representative test case is failure-revealing.

\textbf{Results}. For each OTA update, \Cref{fig:ResultsRQ2TimeFailures} shows the percentage of failure-revealing test cases detected by \crisp for different values of the threshold parameter ($\theta$)  and the corresponding time savings. 
Each point is labeled with the value assigned to the threshold parameter ($\theta$). 
The x-coordinate and the y-coordinate of each point indicate the corresponding savings and the percentage of failure-revealing test cases, respectively.

\Cref{fig:ResultsRQ2TimeFailures} shows that increasing the value assigned to the configuration parameter $\theta$ increases the time savings, since more test cases are considered ``similar''. Therefore, the larger the value assigned to the threshold parameter, the fewer test cases are executed.
The time savings can reach $14305.89s$ ($\approx$4 hours) for OTA1, OTA2 and $10311.72s$ ($\approx$3 hours) for OTA3, OTA4, and OTA5. 
These savings are significant considering that re-running the relevant test cases for OTA1, OTA2, OTA3, OTA4, and OTA5 requires respectively $53690s $ ($\approx$15 hours), $53690s$ ($\approx$15 hours), $34325s$ ($\approx$ 9.5 hours), $34325s$ ($\approx$9.5 hours), and $34325s$ ($\approx$9.5 hours).
Therefore, the time savings reach the 26\%, 26\%, 30\%, 30\%, 30\% for the highest values assigned to the threshold.

\Cref{fig:ResultsRQ2TimeFailures} shows that increasing the value assigned to the configuration parameter $\theta$ decreases the percentage of failures discovered, since by reducing the number of test cases executed after an OTA, the higher is the probability for some failure-revealing test cases not to be executed.
For OTA1, OTA2, OTA3, OTA4, and OTA5, the selection approach can reduce the percentage of failure-revealing test cases discovered by $15.22\%$, $4.35\%$, $18.19\%$, $68\%$, and $0\%$.

Note that the results obtained with the highest threshold value correspond to what we would have obtained by considering higher threshold values. 
Indeed, our test case selection procedure (see \Cref{sec:selection}) ensures that at least one test case for each scenario ($s$) is executed.
The highest threshold values we consider already ensure that exactly one test case (the most critical) for each scenario ($s$) is executed.
Thus, no further improvement is possible with higher thresholds.

Our results indicate the presence of an implicit trade-off between the time savings and the failure-revealing capabilities that depends on the value assigned to the threshold parameter.
For OTA1, OTA2, OTA3 and OTA4, the percentage of failure-revealing  test cases decreases as the threshold value increases. 
Remarkably, for OTA1 and OTA2, and low threshold values ($\theta < 5$), \crisp successfully detects $100\%$ of the failure-revealing test cases while already achieving a substantial reduction in simulation time ($1.5 \cdot 10^{4}$ s $\approx$ 4 hours) offering time saving without any drawback in terms of capabilities for the identification of the failure-revealing test cases.
As the threshold increases further, the time savings continue to grow, but the percentage of detected failures gradually decreases, i.e., beyond this point, additional time savings are obtained at the expense of reducing the number of failure-revealing test cases that are detected by the proposed solution. 
%
For OTA5, all failure-revealing test cases correspond to the same failure scenario and exhibit very similar behavior. 
As the similarity threshold increases, a larger number of test cases are grouped together and represented by a single representative test case, leading to higher time savings.
Since the selected representative is always the most critical test case and is itself failure-revealing, the failure scenario remains covered regardless of the threshold value. 
Therefore, increasing the threshold does not reduce the percentage of detected failure-revealing test cases, while it significantly decreases the number of tests that need to be executed.

These considerations confirm that the selection of an appropriate value for the threshold parameter ($\theta$) plays a key role in balancing simulation cost reduction and failure detection effectiveness and that engineers can make different decisions depending on their domain knowledge and the type of the update.


\begin{Answer}[RQ2 --- Answer]
	The proposed selection strategy offers significant time savings while maintaining practical capabilities in detecting failure-revealing capabilities. 
    Engineers can configure \crisp by reaching different  trade-off between these two aspects depending on the specific needs of their application.
\end{Answer}

%% file: sections/discussion.tex
\section{Discussion} \label{sec:discussion}

This section presents
 our reflections and lessons learned (\Cref{sec:ll}), and discusses threats to validity (\Cref{sec:threats}).

\subsection{Reflections and Lessons Learned} \label{sec:ll}
We learned the following lessons from our experimentation.

\medskip
\noindent\textbf{L1. Criticality-based information is an effective driver for test case prioritization.}
Our results suggest that criticality-based information, such as the fitness value computed through \Hecate, can be effectively used to guide the prioritization of test cases. 
By ranking test case execution based on their criticality, \crisp outperformed both random ordering and \apdisc, in terms of APFD. 
This finding indicates that test cases exhibiting smaller criticality margins are more likely to reveal failures after an OTA update. 
More broadly, our results suggest that criticality-based measures provide valuable information about the failure-revealing potential of test cases and can therefore serve as an effective basis for prioritization in CPS testing environments.

\noindent\textbf{L2. Criticality-based information is an effective driver for test case selection.}
The results from our empirical evaluation suggest that criticality-based information can effectively guide test case selection for OTA of CPSs. 
Across the considered updates, \crisp improved the efficiency of the testing process by reducing the number of simulations required to identify failure-revealing test cases (if selection is performed). 
These findings indicate that criticality-based information can serve as a practical indicator for identifying test cases that are most relevant for safety analysis after software updates.

\noindent\textbf{L3. The criticality-based threshold value significantly affects the effectiveness of the selection procedure.}
Our results show that the value assigned to the threshold parameter ($\theta$) significantly affects the effectiveness of the test case selection procedure.
The value assigned to this parameter enables engineers to find valuable trade-offs between effectiveness (capability of detecting failure-revealing test cases) and efficiency (time required to find these test cases). 
Specifically, a higher value assigned to the parameter $\theta$ reduces the simulation time, but also decreases the capability of the test case selection procedure in identifying failure-revealing test cases.  
For example, when analyzing the ACC component, setting $\theta$ at the value $5$ significantly reduced the testing time, while not impacting the failure detection rate.
In contrast, for the LKA component, no value of $\theta$ was effective in leaving the failure detection rate unchanged.
This suggests that the optimal threshold value is application-dependent and should be calibrated according to the characteristics of the functionality being tested, as well as to the criticality of the system.

\noindent\textbf{L4. Criticality-based information approaches are more effective when configuration variability has a limited influence on system behavior.}
The effectiveness of our criticality-based selection and prioritization strategy depends on the degree to which different configurations exhibit ``similar'' behavior. 

Our evaluation indicates that \crisp is most effective when two conditions hold. 
First, OTA updates should preserve the relationship between baseline criticality and post-update behavior. 
Large structural modifications (e.g., OTA2) weaken this relationship and reduce the effectiveness of criticality-based prioritization (see RQ1). 
Second, different configurations should exhibit sufficiently similar behavior under the same driving scenario. 
When configuration variability becomes dominant (e.g., OTA4), representative selection becomes less effective because baseline criticality is no longer predictive across configurations.

In the ACC case, in general, vehicle configurations often produced comparable criticality values, enabling substantial reductions in simulation effort with limited impact on failure detection, when test selection is applied. 
In contrast, LKA-related updates exhibited greater sensitivity to differences among vehicles and scenarios, reducing the effectiveness of \crisp in selecting test cases. 
A similar trend emerged for test prioritization. 

Our findings suggest that criticality-based test reduction and prioritization techniques offer larger benefits when it is possible to identify groups of configurations that cluster and contain test cases with similar criticality. 

\noindent\textbf{L5. Prioritization and selection activities are complementary.}
Criticality-based information effectively supported two distinct testing objectives. 
On the one hand, it can be used to prioritize test cases, enabling earlier failure detection without reducing the test suite size. 
On the other hand, it can support test case selection, reducing the number of executed simulations while maintaining an acceptable failure detection rate. 
Our results suggest that prioritization and selection should not be viewed as competing alternatives, but rather as complementary mechanisms that address different constraints encountered during OTA update validation. 
When sufficient computational resources are available, prioritization can accelerate failure discovery while preserving complete coverage. 
Under stricter time budgets, selection can further reduce testing effort by focusing execution on the most informative configurations. 
The selected test cases can be prioritized depending on their criticality, enabling earlier discovery of failure-revealing test cases.

\noindent\textbf{L6. Baseline criticality information can be reused to guide OTA validation.}
Prioritization and selection decisions are derived from the criticality information of test cases computed on the baseline system (before the OTA update is deployed). 
Despite relying exclusively on pre-update information, the resulting ordering and selection strategies are effective in identifying failure-revealing test cases after software modifications, especially when these modifications do not significantly disrupt systems' behavior. 
This suggests that baseline criticality information acts as an effective proxy for estimating the failure-revealing potential of test cases in the updated system.

\noindent\textbf{L7. Different criticality measures offer alternative benefits for test case prioritization.}
Unlike AP-DISC, that uses anti-pattern-based information to prioritize test cases,
\crisp uses a fitness measure computed from requirements by means of HECATE.
From our results we learned that these solutions offer complementary benefits (\Cref{sec:rq1}).
We believe that these results confirm findings from the test case generation problem~\cite{formica2023search}, i.e., combining automatically generated fitness functions (e.g., those generated from requirements) and manually defined fitness functions (e.g., anti-pattern-based) may further improve test case prioritization solution and is a promising area for future research.

\subsection{Threats to Validity} \label{sec:threats} 
Our assumptions and problem definition were conducted on an automotive software product line developed in Simulink and involving two ADAS functionalities.
This is a threat to the \emph{external validity} of our results.
Although these systems are representative of relevant automotive functions, we cannot claim that the same results would necessarily hold for other CPS domains or for other automotive functionalities. Future work should extend the evaluation to additional ADAS components and to different types of CPS. 

The vehicles considered in this work are a threat to the \emph{internal validity} of our results. The fact that we considered a large number (seven) of vehicles and specifications (weight, dimensions, braking capacity, power) provided by their manufacturers mitigates this threat.
Additionally, the absolute amount of saved time depends on the simulation environment, the hardware used to run the experiments, and the execution time of each simulation. Therefore, the exact time savings may vary in different experimental settings. 

%% file: sections/related.tex
\section{Related Work} \label{sec:related}

Regression testing checks for failures after software changes, e.g., OTA updates.
Since executing an entire regression test suite can be too costly, prior work has studied test suite minimization, test case selection, and test case prioritization as complementary strategies to address this problem~\cite{Yoo2012}. 
A comparison between black-box and white-box test case prioritization techniques was conducted by Hénard et al.~\cite{Henard2016}. 
This work found that black-box strategies can be competitive with coverage-based approaches, especially when source-code information is unavailable or expensive to collect.
Empirically evaluated coverage-based prioritization strategies for regression testing were introduced by Rothermel et al.~\cite{Rothermel2001}.  Elbaum et al.~\cite{Elbaum2002} found that prioritization can improve fault detection under different cost and fault models. 
Other work has explored search-based prioritization~\cite{Li2007}, coverage criteria such as MC/DC for high-assurance software~\cite{Jones2001}, and large-scale comparisons of static and dynamic prioritization strategies on modern software systems~\cite{Luo2019}.
Finally, Bombarda et al.~\cite{BOMBARDA2026112645,10675919,10.1145/3579028.3609017} investigated the use of combinatorial testing to select test cases for configurable systems, such as the SPLs we use in this work.
Although these studies provide the foundations for our work, they primarily target conventional software systems or discrete program executions. 
In contrast, our work concerns CPSs modeled in \simulink.

Different approaches have investigated test case prioritization for CPSs modeled in \simulink, considering both white-box and black-box techniques~\cite{Arrieta2026}.
Arrieta et al.~\cite{Arrieta2026} demonstrated that among different techniques, considering an antipattern-based approach (i.e., \apdisc) evaluating signal discontinuity is effective for test case prioritization.
Our solution (\crisp) differs from \apdisc in three main ways. 
First, it targets OTA regression testing, where updates are applied incrementally and where previously computed information can be reused. 
Second, \crisp uses robustness values derived from requirements encoded in Test Assessment blocks, rather than relying on coverage, signal anti-patterns, or input/output similarity. 
Finally, \crisp addresses both prioritization and configurable test selection.
Arrieta et al.~\cite{10.1145/3205455.3205490,ARRIETA2019137} proposed multi-objective black-box test case selection techniques for simulation models. 
Unlike \crisp, which uses test case criticality to select test cases, the approach introduced by Arrieta et al. uses objectives such as execution cost and test quality to reduce testing effort.

In the context of autonomous vehicles, Birchler et al.~\cite{birchler2023sdcprioritizer} proposed single- and multi-objective prioritization techniques for self-driving-car simulations, using static road features such as road geometry to order virtual tests.
Unlike the approaches introduced by Birchler et al.~\cite{birchler2023sdcprioritizer}, \crisp does not use any information about the scenario in which the simulation will be performed. 
This makes our approach criticality-driven and useful to perform regression testing after the introduction of an OTA update.

Similar approaches are used to generate test cases for autonomous driving vehicles. 
Kim et al.~\cite{kim2022drivefuzz} generate autonomous-driving tests using driving-quality feedback to expose safety violations. 
Likewise, Cheng et al.~\cite{cheng2023behavexplor} proposed BehAVExplor, which uses behavior diversity to guide testing of autonomous-driving systems.
The objective of these works differs from ours. 
They aim to generate test cases to discover failures. 
In contrast, \crisp assumes that engineers already have a regression test suite for their CPS and focuses on efficiently re-executing that suite after OTA updates.
Other research has shown that variations in vehicle characteristics, such as mass, tire friction, or braking capability, can significantly affect the safety of autonomous driving systems~\cite{pan2023safevar}.
Like Pan et al.~\cite{pan2023safevar}, we consider a case study with different vehicles and characteristics: OTA updates can change their behavior. 

OTA updates have been studied from the perspective of safety assurance, change-impact analysis, and product-line reasoning.
Raeisdanaei et al.~\cite{10.1145/3652620.3687815} propose assurance-case support for evaluating automotive OTA updates, while Murphy et al.~\cite{10.1145/3796233} analyzed the effectiveness of assurance cases for evolving medical devices;
Annable et al.~\cite{10.1007/978-3-031-68606-1_9} study comprehensive change-impact analysis for advanced automotive systems;
Bisceglia et al.~\cite{bisceglia2026safety} propose a contract-driven approach for safety analysis of OTA updates in CPS product lines.
These works address the safety argumentation and impact-analysis problems raised by frequent updates in configurable CPSs. 
\crisp is complementary: rather than constructing or updating the safety argument itself, it supports the testing activity required after an OTA update by prioritizing and selecting the simulations most likely to reveal safety violations early.

Our approach builds on robustness-guided falsification and requirement-driven testing.
Quantitative robustness semantics for temporal properties were introduced to measure how far a signal is from satisfying or violating a requirement~\cite{Donze2010}.
Tools such as S-TaLiRo~\cite{Annpureddy2011}, Breach~\cite{Donze2010b}, \Hecate~\cite{10374027}, and \ATheNA~\cite{formica2023search} use robustness values to guide falsification of CPSs. 
\crisp adapts this idea to regression testing after OTA updates: it does not use robustness to generate new tests, but to rank and select existing tests according to their baseline criticality.

Summarizing, unlike existing work from the literature, in this work (a) our approach explicitly targets the prioritization testing problem after the introduction of an OTA update,  (b) we use fitness information derived from Test Assessment as a proxy of test case criticality to prioritize simulations, and (c) we combine the ranking given by test cases criticality with a selection strategy that reduces the number of simulations to be executed while preserving a high fault detection capability. 

%% file: sections/conclusions.tex
\section{Conclusion} \label{sec:conclusion}
This paper proposed a criticality-driven approach to support the safety analysis of OTA updates for CPS. 
Re-executing the full test suite after an OTA update can be highly time consuming and impractical. 
We proposed \crisp, a criticality-based solution that solves this problem and consists of two complementary techniques, both grounded in the concept of test case criticality. 

The first contribution is a test case prioritization technique that uses criticality information to rank test cases from most to least critical. The underlying idea is that test cases closer to violating the test assessment conditions (which encode system requirements), are more likely to reveal failures following an OTA update and should therefore be executed first.

The second contribution is a test case selection technique that exploits behavioral similarities among different vehicles within the same scenario. 
When the behavioral difference of different vehicles is small, only the most critical test case needs to be executed, with its outcome serving as a a proxy for the remaining test cases. 

To evaluate \crisp, we created a large and representative automotive software product line developed in Simulink. The case study encompasses multiple vehicles, scenarios and  software components, yielding a product line of 224 products used as test cases. Re-executing all these test cases requires 59 hours of simulation time. 
We further considered five different OTA updates: two affecting the ACC and three affecting the LKA. 

Our results confirm the effectiveness of \crisp. It  detects failures earlier than both random and \apdisc baseline prioritization approaches, and reduces the total simulation time needed to test a new update while preserving a high failure detection capability. 
For ACC updates, conservative threshold values allow the approach to save simulation time while still detecting all failures. 
For LKA updates, \crisp continues to reduce the simulation time needed, though with a more pronounced reduction in failure detection capability. 
Overall, our evaluation demonstrates that \crisp achieves practical trade-offs between failure detection capability and simulation time savings.